\documentstyle[10pt,epsf,epsfig,hangcaption,xspace,amssymb,amsfonts,amsmath,amsthm,cite,
dp_delphititle,lineno]{dp_delphi}
\makeindex
\def\DpPaperGroup{EP}
\def\DpPaperRef{2003-056}
\def\DpDate{1 April 2003}
\def\DpAuthors{DELPHI Collaboration}
\def\DpTitle{Measurement of the ${\bold \Lambda_b^0}$ Decay Form Factor}
\def\DpSubmit{(Phys. Lett. B585 (2004) 63-84)}
\def\DpComment{ }
\def\DpEMail{ }
\include{newcom}
\newcommand{\lc}{\Lambda_c}
\newcommand{\mlc}{m_{\Lambda_c}}
\newcommand{\mlb}{m_{\Lambda_b}}
\newcommand{\ro}{{\hat{\rho}}^2}
\begin{document}
%%%%%%%%%%%%%%%%%%%%%%%%%% They are a problem with Coll.Sty ?
\makeatletter
%\input{dp_system:coll.sty}
% Collapse citation numbers to ranges.  Non-numeric and undefined labels
% are handled.  No sorting is done.  E.g., 1,3,2,3,4,5,foo,1,2,3,?,4,5
% gives 1,3,2-5,foo,1-3,?,4,5
\newcount\@tempcntc
\def\@citex[#1]#2{\if@filesw\immediate\write\@auxout{\string\citation{#2}}\fi
  \@tempcnta\z@\@tempcntb\m@ne\def\@citea{}\@cite{\@for\@citeb:=#2\do
    {\@ifundefined
       {b@\@citeb}{\@citeo\@tempcntb\m@ne\@citea\def\@citea{,}{\bf ?}\@warning
       {Citation `\@citeb' on page \thepage \space undefined}}%
    {\setbox\z@\hbox{\global\@tempcntc0\csname b@\@citeb\endcsname\relax}%
     \ifnum\@tempcntc=\z@ \@citeo\@tempcntb\m@ne
       \@citea\def\@citea{,}\hbox{\csname b@\@citeb\endcsname}%
     \else
      \advance\@tempcntb\@ne
      \ifnum\@tempcntb=\@tempcntc
      \else\advance\@tempcntb\m@ne\@citeo
      \@tempcnta\@tempcntc\@tempcntb\@tempcntc\fi\fi}}\@citeo}{#1}}
\def\@citeo{\ifnum\@tempcnta>\@tempcntb\else\@citea\def\@citea{,}%
  \ifnum\@tempcnta=\@tempcntb\the\@tempcnta\else
   {\advance\@tempcnta\@ne\ifnum\@tempcnta=\@tempcntb \else \def\@citea{--}\fi
    \advance\@tempcnta\m@ne\the\@tempcnta\@citea\the\@tempcntb}\fi\fi}
 
\makeatother
%%%%%%%%%%%%%%%%%%%%%%%%%% ??????????????????????????????????
% Generate the title page
\begin{titlepage}
\pagenumbering{roman}
\CERNpreprint{\DpPaperGroup}{\DpPaperRef} % Reference of the paper
\date{{\small\DpDate}} % Date of the paper
\title{\DpTitle} % Title of the paper
\address{\DpAuthors} % General name of the author(s)
\begin{shortabs} % Start the abstract
\noindent
%===================> Abstract     =====> To be filled <=====%
The form factor of $\Lambda_b^0$ baryons is estimated using $3.46\cdot10^6$ 
hadronic $Z$ decays collected by the DELPHI experiment between 1992 and 1995. 
Charmed $\lc^+$ baryons fully reconstructed in the 
%DB $p K^- \pi^+$, $p \bar{K}^0_S$, and $\Lambda \pi^+ \pi^+ \pi^-$
$p K^- \pi^+$, $p K^0_S$, and $\Lambda \pi^+ \pi^+ \pi^-$
modes, are associated to a lepton with opposite charge 
in order to select $\Lambda_b^0 \to \Lambda_c^+ l^- \overline{\nu}_l$ decays.
From a combined 
likelihood and event rate fit to the distribution 
of the Isgur-Wise variable $w$, and using the 
Heavy Quark Effective Theory (HQET), the slope of the $b$-baryon 
form factor is measured to be~: 
$$\ro = 2.03 \pm 0.46 \;({\mathrm stat}) ^{+0.72}_{-1.00} \;({\mathrm syst}). $$
The exclusive semileptonic branching fraction 
$Br(\Lambda_b^0 \to \Lambda_c^+ l^- \bar{\nu}_l)$ can be derived 
from $\ro$ and is found to be  
%DB $(5.0 ^{+1.8}_{-1.4} \;({\mathrm stat+syst})) \%$.
$(5.0^{+1.1}_{-0.8} \;\mbox{(stat)} ^{+1.6}_{-1.2} \;\mbox{(syst)}) \%$.
Limits on other branching fractions are also obtained. 

\end{shortabs}
\vfill
\begin{center}
\DpSubmit \ \\ % Horrible hack to allow to have DpSubmit empty
\DpComment \ \\
\DpEMail \ \\
\end{center}
\vfill
\clearpage
\headsep 10.0pt
\addtolength{\textheight}{10mm}
\addtolength{\footskip}{-5mm}
\begingroup
% Commands to process the author names
%
\newcommand{\DpName}[2]{\hbox{#1$^{\ref{#2}}$},\hfill}
\newcommand{\DpNameTwo}[3]{\hbox{#1$^{\ref{#2},\ref{#3}}$},\hfill}
\newcommand{\DpNameThree}[4]{\hbox{#1$^{\ref{#2},\ref{#3},\ref{#4}}$},\hfill}
\newskip\Bigfill \Bigfill = 0pt plus 1000fill
\newcommand{\DpNameLast}[2]{\hbox{#1$^{\ref{#2}}$}\hspace{\Bigfill}}
%
%\small
\footnotesize
\noindent
\DpName{J.Abdallah}{LPNHE}
\DpName{P.Abreu}{LIP}
\DpName{W.Adam}{VIENNA}
\DpName{P.Adzic}{DEMOKRITOS}
\DpName{T.Albrecht}{KARLSRUHE}
\DpName{T.Alderweireld}{AIM}
\DpName{R.Alemany-Fernandez}{CERN}
\DpName{T.Allmendinger}{KARLSRUHE}
\DpName{P.P.Allport}{LIVERPOOL}
\DpName{U.Amaldi}{MILANO2}
% ...
\normalsize
\endgroup
\titlefoot{Department of Physics and Astronomy, Iowa State
     University, Ames IA 50011-3160, USA
    \label{AMES}}
\titlefoot{Physics Department, Universiteit Antwerpen,
     Universiteitsplein 1, B-2610 Antwerpen, Belgium \\
     \indent~~and IIHE, ULB-VUB,
     Pleinlaan 2, B-1050 Brussels, Belgium \\
     \indent~~and Facult\'e des Sciences,
     Univ. de l'Etat Mons, Av. Maistriau 19, B-7000 Mons, Belgium
    \label{AIM}}
\titlefoot{Physics Laboratory, University of Athens, Solonos Str.
     104, GR-10680 Athens, Greece
    \label{ATHENS}}
\titlefoot{Department of Physics, University of Bergen,
     All\'egaten 55, NO-5007 Bergen, Norway
    \label{BERGEN}}
\titlefoot{Dipartimento di Fisica, Universit\`a di Bologna and INFN,
     Via Irnerio 46, IT-40126 Bologna, Italy
    \label{BOLOGNA}}
\titlefoot{Centro Brasileiro de Pesquisas F\'{\i}sicas, rua Xavier Sigaud 150,
     BR-22290 Rio de Janeiro, Brazil \\
     \indent~~and Depto. de F\'{\i}sica, Pont. Univ. Cat\'olica,
     C.P. 38071 BR-22453 Rio de Janeiro, Brazil \\
     \indent~~and Inst. de F\'{\i}sica, Univ. Estadual do Rio de Janeiro,
     rua S\~{a}o Francisco Xavier 524, Rio de Janeiro, Brazil
    \label{BRASIL}}
\titlefoot{Coll\`ege de France, Lab. de Physique Corpusculaire, IN2P3-CNRS,
     FR-75231 Paris Cedex 05, France
    \label{CDF}}
\titlefoot{CERN, CH-1211 Geneva 23, Switzerland
    \label{CERN}}
\titlefoot{Institut de Recherches Subatomiques, IN2P3 - CNRS/ULP - BP20,
     FR-67037 Strasbourg Cedex, France
    \label{CRN}}
\titlefoot{Now at DESY-Zeuthen, Platanenallee 6, D-15735 Zeuthen, Germany
    \label{DESY}}
\titlefoot{Institute of Nuclear Physics, N.C.S.R. Demokritos,
     P.O. Box 60228, GR-15310 Athens, Greece
    \label{DEMOKRITOS}}
\titlefoot{FZU, Inst. of Phys. of the C.A.S. High Energy Physics Division,
     Na Slovance 2, CZ-180 40, Praha 8, Czech Republic
    \label{FZU}}
\titlefoot{Dipartimento di Fisica, Universit\`a di Genova and INFN,
     Via Dodecaneso 33, IT-16146 Genova, Italy
    \label{GENOVA}}
\titlefoot{Institut des Sciences Nucl\'eaires, IN2P3-CNRS, Universit\'e
     de Grenoble 1, FR-38026 Grenoble Cedex, France
    \label{GRENOBLE}}
\titlefoot{Helsinki Institute of Physics, P.O. Box 64,
     FIN-00014 University of Helsinki, Finland
    \label{HELSINKI}}
\titlefoot{Joint Institute for Nuclear Research, Dubna, Head Post
     Office, P.O. Box 79, RU-101 000 Moscow, Russian Federation
    \label{JINR}}
\titlefoot{Institut f\"ur Experimentelle Kernphysik,
     Universit\"at Karlsruhe, Postfach 6980, DE-76128 Karlsruhe,
     Germany
    \label{KARLSRUHE}}
\titlefoot{Institute of Nuclear Physics,Ul. Kawiory 26a,
     PL-30055 Krakow, Poland
    \label{KRAKOW1}}
\titlefoot{Faculty of Physics and Nuclear Techniques, University of Mining
     and Metallurgy, PL-30055 Krakow, Poland
    \label{KRAKOW2}}
\titlefoot{Universit\'e de Paris-Sud, Lab. de l'Acc\'el\'erateur
     Lin\'eaire, IN2P3-CNRS, B\^{a}t. 200, FR-91405 Orsay Cedex, France
    \label{LAL}}
\titlefoot{School of Physics and Chemistry, University of Lancaster,
     Lancaster LA1 4YB, UK
    \label{LANCASTER}}
\titlefoot{LIP, IST, FCUL - Av. Elias Garcia, 14-$1^{o}$,
     PT-1000 Lisboa Codex, Portugal
    \label{LIP}}
\titlefoot{Department of Physics, University of Liverpool, P.O.
     Box 147, Liverpool L69 3BX, UK
    \label{LIVERPOOL}}
\titlefoot{Dept. of Physics and Astronomy, Kelvin Building,
     University of Glasgow, Glasgow G12 8QQ
    \label{GLASGOW}}
\titlefoot{LPNHE, IN2P3-CNRS, Univ.~Paris VI et VII, Tour 33 (RdC),
     4 place Jussieu, FR-75252 Paris Cedex 05, France
    \label{LPNHE}}
\titlefoot{Department of Physics, University of Lund,
     S\"olvegatan 14, SE-223 63 Lund, Sweden
    \label{LUND}}
\titlefoot{Universit\'e Claude Bernard de Lyon, IPNL, IN2P3-CNRS,
     FR-69622 Villeurbanne Cedex, France
    \label{LYON}}
\titlefoot{Dipartimento di Fisica, Universit\`a di Milano and INFN-MILANO,
     Via Celoria 16, IT-20133 Milan, Italy
    \label{MILANO}}
\titlefoot{Dipartimento di Fisica, Univ. di Milano-Bicocca and
     INFN-MILANO, Piazza della Scienza 2, IT-20126 Milan, Italy
    \label{MILANO2}}
\titlefoot{IPNP of MFF, Charles Univ., Areal MFF,
     V Holesovickach 2, CZ-180 00, Praha 8, Czech Republic
    \label{NC}}
\titlefoot{NIKHEF, Postbus 41882, NL-1009 DB
     Amsterdam, The Netherlands
    \label{NIKHEF}}
\titlefoot{National Technical University, Physics Department,
     Zografou Campus, GR-15773 Athens, Greece
    \label{NTU-ATHENS}}
\titlefoot{Physics Department, University of Oslo, Blindern,
     NO-0316 Oslo, Norway
    \label{OSLO}}
\titlefoot{Dpto. Fisica, Univ. Oviedo, Avda. Calvo Sotelo
     s/n, ES-33007 Oviedo, Spain
    \label{OVIEDO}}
\titlefoot{Department of Physics, University of Oxford,
     Keble Road, Oxford OX1 3RH, UK
    \label{OXFORD}}
\titlefoot{Dipartimento di Fisica, Universit\`a di Padova and
     INFN, Via Marzolo 8, IT-35131 Padua, Italy
    \label{PADOVA}}
\titlefoot{Rutherford Appleton Laboratory, Chilton, Didcot
     OX11 OQX, UK
    \label{RAL}}
\titlefoot{Dipartimento di Fisica, Universit\`a di Roma II and
     INFN, Tor Vergata, IT-00173 Rome, Italy
    \label{ROMA2}}
\titlefoot{Dipartimento di Fisica, Universit\`a di Roma III and
     INFN, Via della Vasca Navale 84, IT-00146 Rome, Italy
    \label{ROMA3}}
\titlefoot{DAPNIA/Service de Physique des Particules,
     CEA-Saclay, FR-91191 Gif-sur-Yvette Cedex, France
    \label{SACLAY}}
\titlefoot{Instituto de Fisica de Cantabria (CSIC-UC), Avda.
     los Castros s/n, ES-39006 Santander, Spain
    \label{SANTANDER}}
\titlefoot{Inst. for High Energy Physics, Serpukov
     P.O. Box 35, Protvino, (Moscow Region), Russian Federation
    \label{SERPUKHOV}}
\titlefoot{J. Stefan Institute, Jamova 39, SI-1000 Ljubljana, Slovenia
     and Laboratory for Astroparticle Physics,\\
     \indent~~Nova Gorica Polytechnic, Kostanjeviska 16a, SI-5000 Nova Gorica, Slovenia, \\
     \indent~~and Department of Physics, University of Ljubljana,
     SI-1000 Ljubljana, Slovenia
    \label{SLOVENIJA}}
\titlefoot{Fysikum, Stockholm University,
     Box 6730, SE-113 85 Stockholm, Sweden
    \label{STOCKHOLM}}
\titlefoot{Dipartimento di Fisica Sperimentale, Universit\`a di
     Torino and INFN, Via P. Giuria 1, IT-10125 Turin, Italy
    \label{TORINO}}
\titlefoot{INFN,Sezione di Torino, and Dipartimento di Fisica Teorica,
     Universit\`a di Torino, Via P. Giuria 1,\\
     \indent~~IT-10125 Turin, Italy
    \label{TORINOTH}}
\titlefoot{Dipartimento di Fisica, Universit\`a di Trieste and
     INFN, Via A. Valerio 2, IT-34127 Trieste, Italy \\
     \indent~~and Istituto di Fisica, Universit\`a di Udine,
     IT-33100 Udine, Italy
    \label{TU}}
\titlefoot{Univ. Federal do Rio de Janeiro, C.P. 68528
     Cidade Univ., Ilha do Fund\~ao
     BR-21945-970 Rio de Janeiro, Brazil
    \label{UFRJ}}
\titlefoot{Department of Radiation Sciences, University of
     Uppsala, P.O. Box 535, SE-751 21 Uppsala, Sweden
    \label{UPPSALA}}
\titlefoot{IFIC, Valencia-CSIC, and D.F.A.M.N., U. de Valencia,
     Avda. Dr. Moliner 50, ES-46100 Burjassot (Valencia), Spain
    \label{VALENCIA}}
\titlefoot{Institut f\"ur Hochenergiephysik, \"Osterr. Akad.
     d. Wissensch., Nikolsdorfergasse 18, AT-1050 Vienna, Austria
    \label{VIENNA}}
\titlefoot{Inst. Nuclear Studies and University of Warsaw, Ul.
     Hoza 69, PL-00681 Warsaw, Poland
    \label{WARSZAWA}}
\titlefoot{Fachbereich Physik, University of Wuppertal, Postfach
     100 127, DE-42097 Wuppertal, Germany \\
\noindent
{$^\dagger$~deceased}
    \label{WUPPERTAL}}
\addtolength{\textheight}{-10mm}
\addtolength{\footskip}{5mm}
\clearpage
\headsep 30.0pt
\end{titlepage}
%%%%%%%%%%%%%%%%%%%%%%%%%
%
% Change for the document body
%%\pagestyle{heading} % for page numbering
\pagenumbering{arabic} % page numbering in number
\setcounter{footnote}{0} %
\large
%\linenumbers %%%CD
%*****************************************************************************
\section{Introduction}

The knowledge of the $B$-meson form factor has recently improved 
thanks to a wealth of new experimental results, as reported for 
instance in 
references~\cite{ref:delphivcb1,ref:alephvcb,ref:opalvcb,ref:delphivcb2,ref:cleovcb,ref:bellevcb}.
Semileptonic decays of $B$ mesons into $D$ and $D^{\ast}$ final 
states can be understood in the context of the Heavy Quark Effective Theory
(HQET), where the four form factors remaining when the lepton mass
is neglected can be expressed in terms of a single 
Isgur-Wise function $\xi_M$~\footnote{The subscript $M$ has been added to 
indicate that this form factor 
applies to meson decay, while the subscript $B$ will be used for baryons.},
which will be defined in section~2. 

In this paper, the semileptonic $b$-baryon 
decay $\Lambda_b^0 \to \Lambda_c^+ l^- \bar{\nu}_l$ 
(with $l^- = e^-$  or $\mu^-$) is investigated~\footnote{The notations 
$\Lambda_b^0$ and $\Lambda_c^+$ will implicitly 
stand for both the baryon and antibaryon, with the proper
inversion of the signs of the lepton and of the $\Lambda_c^+$ decay products.},
where the $\Lambda_c^+$ is fully  reconstructed from its  
decay modes into $p K^- \pi^+$, 
$\Lambda \pi^+ \pi^+\pi^-$, and  $p K^0_S$.
The heavy quark symmetry relates form factors of the 
transition $\Lambda_b^0 \to \Lambda_c^+ l^- \bar{\nu}_l$ to a new single
Isgur-Wise function $\xi_B$, as explained 
in~\cite{ref:Grinstein,ref:Isgur,ref:Georgi} 
and predicts its absolute value when the final 
$\Lambda_c^+$ is at rest in the $\Lambda_b$ frame. 
After a summary of the heavy quark formalism in section~2, 
and a description of the relevant parts of the DELPHI detector
in section~3, the selection of the 
different $\Lambda_c^+$-lepton channels 
candidates is described in section~4. 
The dominant contribution to the $\Lambda_c^+ l^- \bar{\nu}_l$ final 
state is expected to come from the $\Lambda_b^0$ baryon, and the 
relevant contaminations from $B$-mesons, from other $b$-baryons, 
or from other hadronic final states with additional pions
are investigated in sections~5 and~6.
In section~7 a direct determination 
of the semileptonic branching fraction is presented, while  
the fit to the distribution of the Isgur-Wise variable 
$w$ is described in section~8. The semileptonic branching fraction
and the $w$ distribution are then combined into a single  
measurement of the slope parameter $\ro$ of the $b$-baryon form factor, 
assuming the validity of HQET predictions.  

\section{The semileptonic decay form factor of ${\bold b}$-baryons}

A complete description of the form factor formalism in semileptonic
decays and theoretical predictions 
can be found in~\cite{ref:neubert,ref:lattice,ref:sumrules}.
The form factors are functions of the four-momentum transfer $q^2$ in the 
transition, with $q^2 = (p_{l} + p_{\bar{\nu}_l})^2$
where $p_{l}$ and $p_{\bar{\nu}_l}$ are the charged lepton and neutrino 
four-momenta, respectively.
Isgur and Wise introduce the dimensionless variable $w$, scalar product
of the four-velocities of the $\Lambda_b^0$ and $\Lambda_c^+$: 
\begin{eqnarray}
w = v_{\Lambda_b} \cdot v_{\Lambda_c} = (\mlb^2 + \mlc^2 - q^2) / (2 \mlb \mlc).
\end{eqnarray}

The hadronic current in the weak decay of a beauty baryon 
($J^P = 1/2^+$) to a charmed baryon ($J^P = 1/2^+$), as in 
the transition $\Lambda_b^0 \to \Lambda_c^+ W^-$, involves 
six form factors, three  vectors $F_i$ and three axial-vectors $G_i$.   
In the decay $\Lambda_b^0 \to \Lambda_c^+ l^- \bar{\nu}_l$,
the variable $w$ ranges from 1 (highest transfer, final 
$\Lambda_c^+$ at rest) to a value close to 1.44 (smallest transfer, 
$q^2 = m_l^2$). Among the six form factors which can contribute 
to the semileptonic decay of $J^P = 1/2^+$ baryons, only 
$F_1$ and $G_1$ survive in the limit of infinite $b$ and $c$ quark 
masses, the HQET limit. In addition, they 
are equal and can be expressed in terms of a single function
$\xi_B(w)$.  

The differential decay width of the transition 
$\Lambda_b^0 \to \Lambda_c^+ l^- \bar{\nu}_l$
can be obtained from~\cite{ref:holdom} in the 
approximation where $m_{light}/m_Q$ terms are neglected, 
where $m_{light}$ is the mass of the light quark system 
and $m_Q$ stands for the heavy quark:
\begin{eqnarray}
{{d\Gamma}\over{d w }} \, = \, G \, K(w ) \, \xi_B^2(w),
\end{eqnarray}
where the constant $G$ is: 
\begin{eqnarray}
G =  {{2}\over{3}} {{{G_F}^2}\over{(2\pi)^3}} 
\left |V_{cb} \right|^2 m^4_{\Lambda_b} r^2
\; \; \; \hbox{ with } \; \; \; r = \mlc/\mlb,
\end{eqnarray}
and the kinematical factor $K(w)$ is: 
\begin{eqnarray}
K(w) = P 
\left[
3 w (1 - 2 r w +r^2) + 2 r (w^2 - 1)
\right]
\; \; \hbox{ with } \; \; 
P = \mlc \sqrt{w^2 - 1}.
\end{eqnarray}
The Isgur-Wise function $\xi_B(w)$ will be studied in the
present paper. 
  
In $B$-meson decays, another function $\xi_M(w)$ 
describes the semileptonic transitions 
$\bar{B} \to (D,D^{\ast}) l^- \bar{\nu}_l$, and a
Taylor expansion is usually assumed for this function:
\begin{eqnarray}
\xi_M(w) = \xi_M( 1 ) \left[ 1 - \ro_M (w-1) + {\cal{O}}((w - 1)^2) \right],
\end{eqnarray}
with $\xi_M(1) = 1$ in the HQET limit. The quadratic terms are constrained by 
dispersion relations~\cite{ref:caprini,ref:boyd,ref:chakraverty}.
Taking into account $(m_{light}/m_c)$ corrections 
and perturbative QCD effects~\cite{ref:neubert,ref:lattice,ref:sumrules}, 
the value of $\xi_M(1)$ is modified in the $B$-meson decay channel
to  $\xi_M(1) = 0.91 \pm 0.04$.
Several experimental determinations of the mesonic form factor 
$\xi_M(w)$ have been performed in the channel 
$\bar{B}^0_d  \to D^{\ast +} l^- \bar{\nu}_l$. Different fits are performed 
in~\cite{ref:delphivcb2} which quotes for a constrained quadratic fit:
\begin{eqnarray}
\ro_M = 1.22  \pm 0.14 \;\hbox{(stat)}.
\end{eqnarray}

Another recent determination is given in~\cite{ref:cleovcb}, and
earlier measurements of $\ro_M$ were performed  
by~\cite{ref:delphivcb1,ref:alephvcb,ref:opalvcb}. 

Many different parametrisations have been proposed for the baryonic
function $\xi_B(w)$, as given in references 
~\cite{ref:holdom,ref:holdom1,ref:sadzikowski}. 
The simplest one is chosen which remains positive in the physical $w$ range:
\begin{eqnarray}
\xi_B(w) = \xi_B(1)\times \exp{[-\ro(w - 1)]} \; .
\end{eqnarray}
The linear and quadratic coefficients in the Taylor expansion
of this function $\xi(w)$ are (within errors) in the domain
allowed by the dispersion relation constraints evaluated 
in~\cite{ref:caprini,ref:boyd,ref:chakraverty}.

The flavour independence of QCD implies 
that, as in the usual isospin symmetry, 
$\xi_B(1) = 1$. It is shown in reference~\cite{ref:holdom} that 
the corrections in $m_{light}/m_Q$ to this result vanish 
at first order for baryons (as expected from the general
result of \cite{ref:luke}),  
and remain small at higher order, while this is not true for mesons.  
The relations between the six baryonic form factors and the Isgur-Wise
function $\xi$ are  however slightly 
modified by the small  $m_{light}/m_Q$ corrections
evaluated in~\cite{ref:holdom}. The perturbative QCD corrections are 
smaller than for mesons and will be neglected, as explained
in~\cite{ref:manoharwhite}. 
In sections~7 and~8 of this paper where the observed
exclusive semileptonic branching fraction is used to infer $\ro$, 
the finite mass corrections, as given by~\cite{ref:holdom},
are included into the relation between the Isgur-Wise function $\xi_B(w)$ and
the semileptonic width (equation~2). These corrections are 
evaluated with a $b$ quark mass $m_b = 4.844$~GeV/$c^2$, and a 
charmed quark mass $m_c = 1.35$~GeV/$c^2$, the numerical values chosen 
in \cite{ref:holdom}.

\section{The DELPHI detector and the simulation} 

The DELPHI detector and its performance have been described in detail 
in~\cite{ref:DELPHIdet1,ref:DELPHIdet2}. 
In the barrel region, a set of cylindrical
detectors, with the $z$ coordinate axis    
oriented along the 1.2 T magnetic field (and the beam) direction, allows 
the tracking of charged particles. 
The silicon Vertex Detector (VD), with an intrinsic resolution 
of 7.6 $\mu$m in the plane transverse to the beam axis and 10-30~$\mu$m
along the $z$ axis, consists of 
three layers. The innermost and outer layers were replaced 
by double-sided silicon microstrips for the 1994-1995 data taking period.  
The Inner Detector (ID) extends between radii of 12 cm and 28 cm 
and gives 24 spatial measurements.
The Time Projection Chamber (TPC) provides up to 16 points between 
30~cm and 122~cm. The Outer Detector (OD), at a radius of 197~cm to 206~cm, 
consists of 5 layers of drift cells. In the plane orthogonal to the beam 
direction, the extrapolation accuracy at the primary vertex of hadronic 
charged particles is found to be 
$\sqrt{20^2 + 65^2/p_t^2}$~$\mu$m~\cite{ref:Chabaud}, where $p_t$ (in~GeV/$c$) 
is the momentum of the particle transverse to the beam axis.

The identification of electrons relies on the electromagnetic calorimeter 
in the barrel region (high density projection chamber HPC), with a relative
energy resolution of $6.5\%$ for electrons at a momentum of 45~GeV/$c$.
Within the HPC acceptance, electrons with a momentum above 3~GeV/$c$ are 
identified with an efficiency of 77\%. The probability that a pion 
be misidentified as an electron is below 1\%.
The muon identification relies mainly on the muon chambers. The selection
criteria used in this work ensure an efficiency of 
identification of 77\% for a misidentification probability of 0.8\%.

 The identification of protons and kaons relies on the DELPHI algorithms 
which take into account the information provided by the
Ring Imaging Cherenkov (RICH), and the 
$dE/dx$ in the Time Projection Chamber (TPC). The liquid and gas radiator
signals of the RICH are used when they were present, mostly in 
the 1994 and 1995 data samples. A neural network program 
is used for the identification of charged particles in the 1994 data, 
while in the other data samples, a simple combination of the RICH
and TPC measurements are considered. 
The proton and kaon tracks are required to be in the barrel 
region, with a polar angle to the $z$ axis fulfilling $|\cos{\theta}| < 0.74$. 
The efficiencies associated to proton and kaon identification 
have been obtained from the simulation, and corrected for the 
small differences between the simulation and the data.
A check of the efficiency estimate was performed on dedicated samples  
of real data, namely $\Lambda$ (for the proton) and reconstructed 
$D^{*\pm}$ in the $K\pi\pi$ channel (for the kaon): 
a good agreement was found with the simulation in the whole momentum range. 
The overall proton  identification efficiencies 
are $(24 \pm 4)\%$ (1992 data), $(21.5 \pm 4)\%$ (1993 data), when
the liquid RICH was not operating, and  
$(42 \pm 2)\%$ in 1994 and 1995.
The fraction of pions misidentified as protons is approximately 
5\% above 3 GeV$/c$ for good operating conditions of the RICH. 

Special samples of events for each potential source of 
$\Lambda_c^+$-lepton final states were generated
using the JETSET 7.3 Parton Shower
program~\cite{ref:Sjostrand}, with a $\Lambda_b^0$ lifetime set to 
1.6~ps. The generated events were followed through the detailed
detector simulation DELSIM. The background events without 
a true $\Lambda_c$ were selected from the general DELPHI 
simulated sample of $q \bar{q}$ events at the $Z$. 
The same sample was used to estimate the fake lepton background.
These events were then processed 
through the same analysis chain as the real data. 
A reweighting of the simulated events, 
which were generated with a constant value $\xi_B(w) = 1 $ 
of the form factor, allows 
the observed distributions to be predicted for all variables, 
and the slope parameter $\ro$ to be tuned in order 
to reproduce the data. A reweighting was also applied
to match the measured $\Lambda_b^0$ lifetime of 1.23~ps~\cite{ref:pdg}.

\section{Event selection}

\subsection{The sample of hadronic events}

Hadronic $Z$ decays collected between 1992 and 1995 were used. 
The centre-of-mass energy was required to lie within 2 GeV of the $Z$ mass. 
Charged particles were required to have a measured momentum between 
0.1~GeV/$c$ and 50~GeV/$c$, a relative error on momentum less 
than 100\%, a track length larger than 30 cm, and a distance 
of closest approach to the interaction point smaller than 5 cm 
radially, and smaller than 10 cm along the beam axis. Neutral particles were 
required to have an energy between 1 GeV and 30 GeV, and a polar angle 
between 20 and 160 degrees. They were assigned a zero mass. 

Hadronic events were then selected using the previous set of 
charged particles with a momentum above 0.4~GeV/$c$. 
Five or more charged particles were required, 
carrying a total energy (assuming them to be pions) 
of more than $0.12 \sqrt{s}$.
A total of 3.46 million events has been obtained.
In the following sections, events will be selected from this sample 
which contain candidates for both a $\Lambda_c$ and a lepton.  
 
\subsection{Lepton selection and ${\bold b}$-tagging}

\begin{figure}[h]
\begin{center}
%DB \epsfxsize = 14 true cm
\epsfxsize = 13 true cm
\epsffile{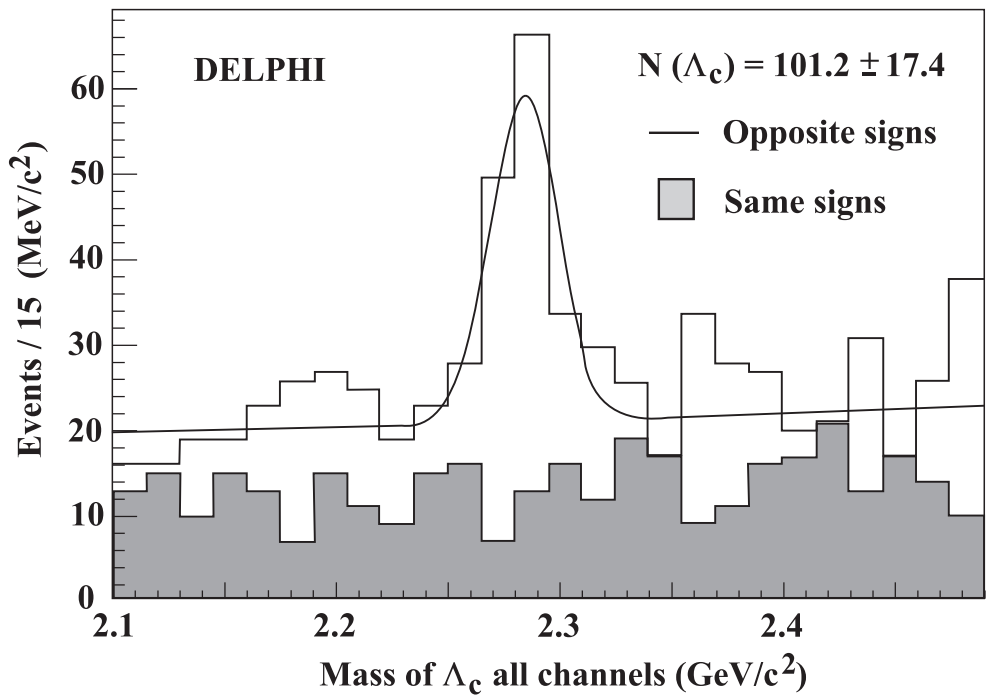}
\caption{\it The $\Lambda_c^+$ mass distribution including 
the decay channels $p K^- \pi^+$, $\Lambda \pi^+\pi^+\pi^-$ and 
$p K^0_S$: 
$\Lambda_c^+ l^-$ combinations (white histogram),
$\Lambda_c^+ l^+$ combinations (hatched histogram).
The curve is the sum of the fitted functions for each channel 
(see section~4.6).}
\label{fig:1}
\end{center}
\end{figure}

The lepton candidates had to satisfy the appropriate identification 
criteria, and a method relying on a neural network 
was used for the electron identification.  
The charged leptons were required to have 
a momentum larger than 3~GeV/$c$ and a 
transverse momentum $p_t$ with respect to the $\Lambda_c^+$ candidates 
(defined in sections~4.3 to~4.5 below) larger than 0.6~GeV/$c$.  
Additional criteria were then introduced to purify the $\Lambda_b^0$ sample:

\begin{itemize}
\item The mass of the $\Lambda_c^+$-lepton system was required to be 
      larger than 3~GeV/$c^2$ (for electrons), or 3.2~GeV/$c^2$ (for muons).
      This selection reduces the potential contributions 
      from semileptonic $B$-meson decays with baryons in the 
      final state and from the $\Lambda_c^+ X l^- \bar{\nu}_l$ final
      states of $\Lambda_b^0$ decays. The lower momentum selection for the
      electron preserves the efficiency, given a lower resolution 
      in this channel.   

\item The probability that all tracks in the event come from the primary  
      vertex was required  to be less than 10\%, 
      as described in~\cite{ref:DELPHIdet2}. The events were then 
      considered as $b$-flavour candidates. This choice is the result of a  
      compromise between the signal yield and the level of 
      combinatorial background in the $\Lambda_c^+$ mass window; 
      it gives a $b$-tagging efficiency of 80\% with a purity of 54\%.   

\item The sign of the lepton charge must be opposite to the 
      $\Lambda_c^+$ charge.
      It can be seen in Figure~\ref{fig:1} that there is no evidence for 
      a $\Lambda_c^+$ signal 
      in the wrong charge mass distribution. The upper limit measured
      from the data using the wrong sign lepton rate is 
      7\% (95\% C.L.) of the right sign sample. Similar limits have been 
      obtained from a simulated sample of $b \bar{b}$ events:
      a fit to the $\Lambda_c^+$ mass distribution 
      of $\Lambda_c^+ l^-$ candidates, as in the actual data analysis,
      gave 438 $\Lambda_c^+ l^-$ combinations. A sample of 
      $13 \pm 5$ $\Lambda_c^+$ candidates were associated 
      to fake leptons. Their fraction is then $(4.3 \pm 1.5)$\% of 
      the signal~\footnote{This number has been corrected for the 
      different rate of fake leptons in the simulation and in the real 
      data, and it has been subtracted from the amount of candidate 
      events in the estimates of absolute rates.}.  
\end{itemize}  
All final states studied in the following must have a track satisfying
the lepton selection requirements.

\subsection{${\bold \Lambda_c^+ \to p K^- \pi^+}$ selection}

Triplets of charged particles of total charge unity, each track with 
at least one hit in the microvertex detector were selected.
The momenta were required to be larger 
than 3~GeV/$c$ (proton candidate), 2~GeV/$c$ (kaon candidate) 
or 1~GeV/$c$ (pion candidate), 
and the total momentum to be larger than 8~GeV/$c$. 
The mass of the $\Lambda_c^+$ candidate had to lie in the 
2.1-2.49~GeV/$c^2$ range. A secondary vertex was fitted with these 
three tracks, requiring a $\chi^2$ 
probability larger than 0.001. The primary vertex was found iteratively
using initially all the tracks of the event and following the procedure 
used in~\cite{ref:delphivertex}. The lepton track was not a priori excluded
from the general vertex fit to avoid a possible bias on the lepton side.
The combined $\Lambda_b^0-\Lambda_c^+$ flight distance was 
then computed as the difference between the secondary and 
the primary vertex. 
It was signed with respect to the momentum direction of the triplet, 
and its projection on the plane transverse to the
beam axis had to be larger than +0.02 cm.  
The particles compatible with both $p$ and $K$ identifications 
were kept, and no identification was applied to the pion candidate.   
The reconstruction efficiency for this channel is 8\%.

\begin{figure}[htb]
\begin{center}
%DB \epsfxsize = 12 true cm
\epsfxsize = 14 true cm
\epsffile{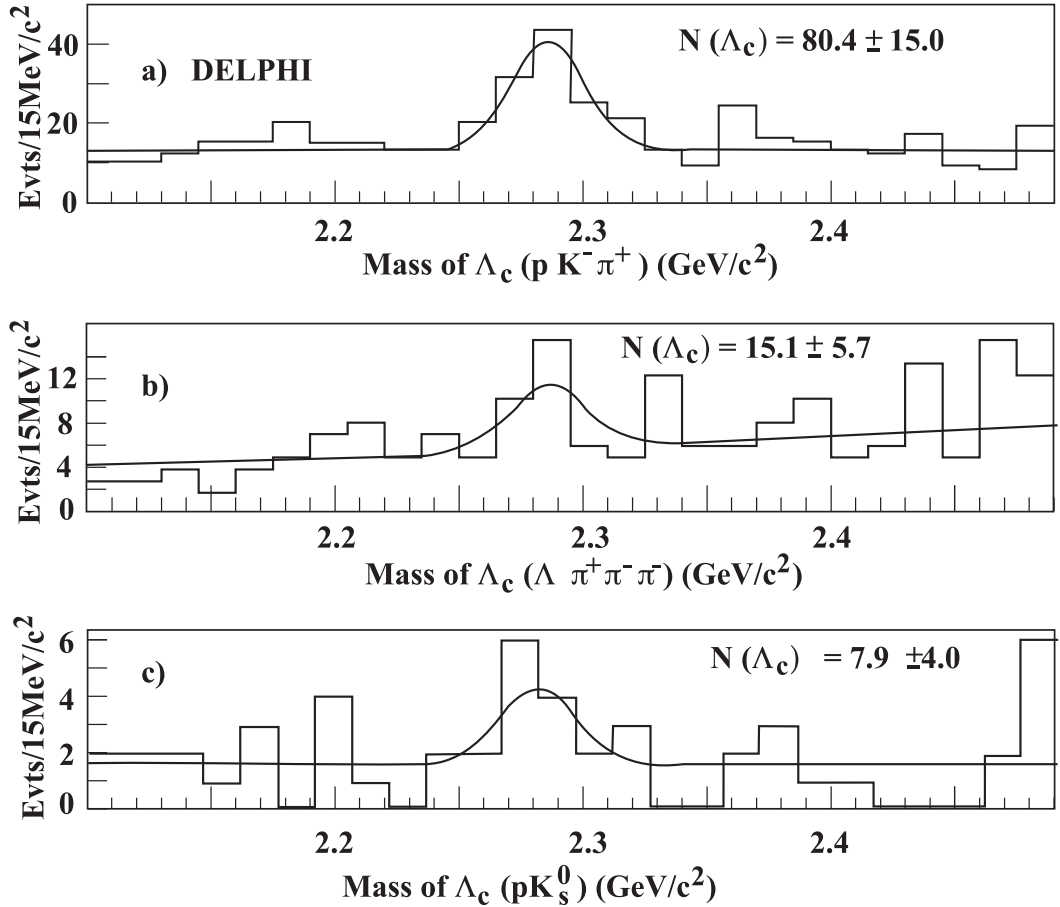}
\caption{\it $\Lambda_c^+$ mass spectrum in the
 a) $p K^- \pi^+$ channel; 
 b) $\Lambda \pi^+ \pi^+ \pi^- $ channel; 
 and c) $p K^0_S$ channel. 
 The fitted functions are described in section~4.3.}
\label{fig:2}
\end{center}
\end{figure}

The invariant mass distribution is shown in Figure~\ref{fig:2}a.
The curves in Figure~\ref{fig:2} are obtained by 
fits of Gaussian distributions in the signal region, added to a 
linear background. The $\Lambda_c$ mass is measured in the 
$p K^- \pi^+$ channel,
and fixed to the same value in the other two. 
The resolution $\sigma$ of the gaussian is left free in the 
$pK^- \pi^+ $ channel, and found to be
13.5 $\pm$ 1.8 ~MeV/$ c^2$. 
The number of $\Lambda_c^+ \to p K^- \pi^+$ events is $80.4 \pm 15.0$.
In the other channels, discussed in the following
sections, the resolution was fixed,
and derived from the previous one according to the ratio of the 
simulated resolutions. After reconstruction in the simulation,
the gaussian resolutions are 
found to be 12~MeV/$c^2$ for the $p K^- \pi^+$ channel, 
15~MeV/$c^2$ for $\Lambda \pi^+\pi^+\pi^-$
and 16~MeV/$c^2$ for $ p K^0_S $.            

\subsection{${\bold \Lambda_c^+ \to \Lambda \pi^+ \pi^+ \pi^-}$ selection}

The $\Lambda$-hyperon candidates were selected by the DELPHI 
algorithm which uses the presence of a remote decay vertex to 
tag the $\Lambda \to p \pi^-$ candidates, as described 
in~\cite{ref:DELPHIdet2}.
The hyperon and $\Lambda_c^+$ momenta were   
required to be larger than 2.5~GeV/$c$ and 10~GeV/$c$, respectively. 
All three pions were required to have 
a momentum larger than 0.4~GeV/$c$, and their tracks to have at least 
one associated hit in the microvertex 
detector. The charge of the triplet had to be 
+1 for $\Lambda$ and -1 for $\overline{\Lambda}$, and 
the three pions had to form a common vertex (the $\Lambda_c^+$ decay vertex) 
with a probability larger than 0.001. The projection of the 
flight distance of the $\Lambda_c^+$ transverse to the beam had to be larger 
than +0.02 cm. 

There is some evidence for a $\Lambda_c^+$ signal in the 
$\Lambda \pi^+ \pi^+ \pi^-$ invariant mass distribution,
as shown in Figure~\ref{fig:2}b, 
with a fitted signal of $15.1 \pm 5.7$ 
$\Lambda_c^+ \to \Lambda \pi^+ \pi^+ \pi^-$ candidates.
All candidate combinations are shown in this figure. Some events
contain more than one accepted combination, and for two events, 
both of them are in the 
$\Lambda_c^+$ mass range $2.260  < M(\Lambda_c^+) < 2.310$~GeV/$c^2$. 

\subsection{${\bold \Lambda_c^+ \to p K^0_S}$ selection}
 
Once a $K^0_S$ candidate had been found~\cite{ref:DELPHIdet2}, 
an identified proton was searched for, such that the
$\Lambda_c$ candidate would belong to the hemisphere defined by the 
lepton track.  
The momenta of the $K^0_S$ and proton had to be larger 
than 3~GeV/$c$.
The ($p K^0_S$) mass distribution is shown in Figure~\ref{fig:2}c, 
with a fitted signal of $7.9 \pm 4.0$ candidates. 

\subsection{The full ${\bold \Lambda_c^+ l^-}$ sample}

The ratios of the number of events obtained in the three channels
are compatible with the known $\Lambda_c^+$ branching 
fractions~\cite{ref:pdg}.
The curve in Figure~\ref{fig:1}, which shows the 
full $\Lambda_c^+ l^-$ sample, 
is obtained from the sum of the three 
Gaussian functions, with their widths as indicated in section~4.3,
and a unique central value. The Gaussian functions are 
weighted by the number of observed events in each channel,
and with a free overall normalisation. The number of $\Lambda_c^+ l^-$
candidates is found to be $101.2 \pm 17.4$.

\section{Other sources of ${\bold \Lambda_c^+ l^- }$ final states}

There are several physical processes which lead to the
$\Lambda_c^+$-lepton final states, in addition to the signal from 
$\Lambda_b^0 \to \Lambda_c^+ l^- \bar{\nu}_l$, named elastic
channel in the following. They arise from
instrumental or physical backgrounds:

\begin{itemize}
\item[a)] the semileptonic decay 
$\Lambda_b^0 \to \Lambda_c^+ \tau^- \bar{\nu}_{\tau}$
with the subsequent decay $\tau^- \to l^- \bar{\nu}_l \nu_{\tau}$;

\item[b)] fake leptons: the fraction of fake leptons, estimated to be 
$(4.3 \pm 1.5)\%$ in section~4.2 will be subtracted from the sample of 
$\Lambda_c^+ l^-$ candidates in sections~7 and~8;

\item[c)] decays of $\bar{B}$ mesons: 
$ \bar{B} \to \Lambda_c^+ \overline{N} l^- \bar{\nu}_l (X)$,
where $\overline{N}$ is an antibaryon and $X$ stands for
any number of $\pi^0$ or ($\pi^+ \pi^-$) pair;

\item[d)] decays into other charmed baryons
$\Lambda_b^0 \to \Lambda_c^{\ast +} l^- \bar{\nu}_l$ , or 
$\Lambda_b^0 \to (\Sigma_c\pi)^+ l^- \bar{\nu}_l$;

\item[e)] non-resonant $\Lambda_c^+ l^- \bar{\nu}_l X$ final states. 
The multiplicity $N_c$ of charged particle tracks compatible with the
$\Lambda_b^0$ vertex, defined as the combined ($\Lambda_cl$) vertex
(ignoring the $\Lambda_c$ lifetime),
the missing mass in the 
$\Lambda_b^0$ decay, and the $\Lambda_c^+ l^-$ mass will be used to 
investigate this component, including as well the decays from d).
The sum of d) and e) final states is named inelastic channels in
the following;

\item[f)]other weakly decaying $b$-baryons, such as 
$\Xi_b$. The production rate of $\Xi_b$ from $b$ quarks is however 
10 times lower than the $\Lambda_b$ production (as discussed in
section~7), and the fraction of $\Lambda_c$ final states from $\Xi_b$, 
if assumed to be similar to that of $\Xi_c$ decays into 
$\Lambda$ baryons, is also less than 10\%. This background has 
thus been neglected.  
\end{itemize}

The final states b) to f)  are background processes,
while the $\tau$ lepton final state a), belongs to the signal.
Its contribution was estimated with the full 
simulation of the decay  
$\Lambda_b^0 \to \Lambda_c^+ \tau^- \bar{\nu}_{\tau}$,
assuming the same couplings as for the other leptons.
The branching fraction is found to be six
times smaller than for $\Lambda_c^+ l^- \bar{\nu}_l$ with light leptons.   
The observed $\tau$ contribution is further suppressed 
by the lepton momentum selections. 
The estimated yield amounts to 2 events in the full 
$\Lambda_c^+$ sample of  101 events and can be neglected.\par 
The number of events from the decay of $\bar{B}$ mesons 
in the $\Lambda_c^+$ 
sample was computed with the full simulation, using the 
value quoted in~\cite{ref:pdg}: 
$Br(\bar{B} \to \Lambda_c^+ \bar{N} l^- \nu_l) < 0.0032$
at the 90\% C.L., which relies on the data from~\cite{ref:CLEO}.
The contamination of such decays into  the sample of 
$\Lambda_c^+ l^- \bar{\nu}_l $ events is found to be less than 
1.5 events (90\% C.L.). The inclusive final states with
an additional charged or neutral pion have a branching fraction
smaller than 0.0064~\cite{ref:pdg}, and their acceptance is 
found from the simulation to be smaller by a factor
0.62, so that they can contribute up to 1.9 events. 
The total contamination of  
$\Lambda_c^+ \bar{N} X l^- \bar{\nu}_l$ events from $\bar{B}$ mesons 
into the $\Lambda_c^+ l^-$ sample 
is thus estimated to be less than 3.4 events (90\% C.L.).  

\section{Charmed baryon contributions to ${\bold b}$-baryon decays}

To extract the form factor corresponding to the 
$\Lambda_b^0 \to \Lambda_c^+ l^- \bar{\nu}_l$ decays 
from the $\Lambda_c^+ l^-$ sample, 
the contributions arising from the elastic and inelastic 
channels must be evaluated. 
Whether resonant charmed hadrons are present 
in the corresponding mass spectra is investigated first: if there is
a large and dominant resonant contribution, its production 
should be described by the appropriate form factors. 
As shown in the following, no signal is observed and the 
corresponding upper limits will be given. 
Other experimental distributions which are
sensitive to the fraction of inelastic channels in the sample
are also considered.
As it will be shown, they can only be understood under 
the assumption of a substantial contribution
from inelastic $\Lambda_c^+ \pi \pi$ final states. 

\subsection{The resonant states}

\begin{figure}
\begin{center}
\epsfxsize = 9.5 true cm
%DB \epsfxsize = 8 true cm
\epsffile{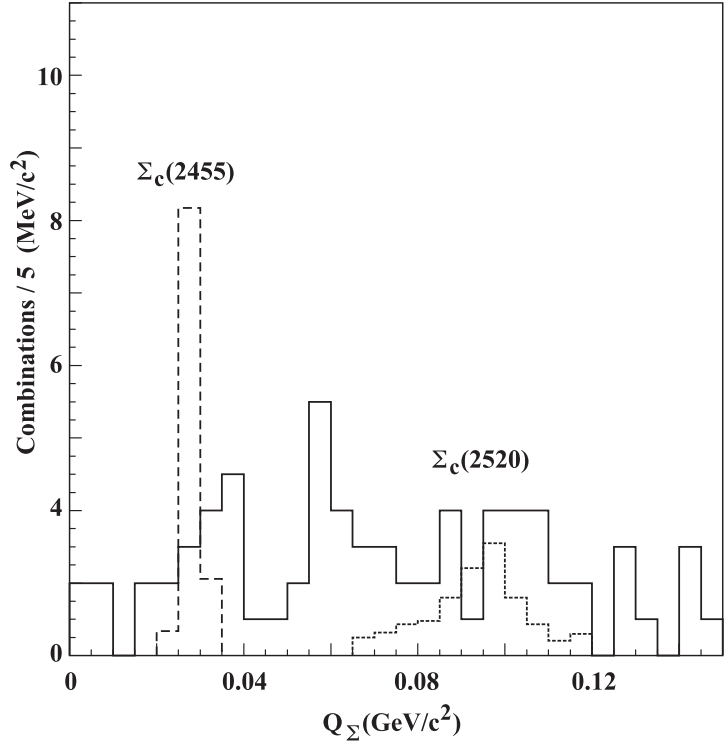}
\caption{\it 
Distribution of 
$Q_{\Sigma} = M(\Lambda_c^+ \pi) - M(\Lambda_c^+)- m_{\pi} $ in data 
(solid line histogram).
The simulated $\Sigma_c$(2455) (dashed line histogram) 
and $\Sigma_c(2520)$ (dotted line histogram) signals each assume
$N_{obs}(\Lambda_b^0 \to \Sigma_c^{++}\pi^- l^- \bar{\nu}_l)$ + 
$N_{obs}(\Lambda_b^0 \to \Sigma_c^0   \pi^+ l^- \bar{\nu}_l) = 
0.1 \; N_{obs}(\Lambda_c^+ (X) l^- \bar{\nu}_l)$.}
\label{fig:3}
\end{center}
\end{figure}

In addition to the elastic $\Lambda_c^+ l^- \bar{\nu}_l$ channel,
several charmed baryon final states can contribute to the
$\Lambda_c^+l^-$ sample, such as 
$(\Sigma_c\pi)^+ X l^- \bar{\nu}_l$, and 
$\Lambda_c^{\ast +} X l^- \bar{\nu}_l$, in which the $\Sigma_c$
and $\Lambda_c^{\ast +}$ decay into a $\Lambda_c^+$. \\
The isospin of the hadronic final state should be $I=0$ within
HQET. In this section, the presence of the HQET-allowed hadronic final states
which correspond to the following decay channels, 
all with a $\Lambda_c^+ \pi \pi$ final state, are investigated:
\begin{itemize}
\item $\Lambda_b^0 \to \Sigma_c^{++}\pi^- l^-\bar{\nu}_l$, with 
      $\Sigma_c^{++} \to \Lambda_c^{+}\pi^+$; 
\item $\Lambda_b^0 \to \Sigma_c^{0}\pi^+ l^-\bar{\nu}_l$, with
      $\Sigma_c^0 \to \Lambda_c^+ \pi^-$;
\item $\Lambda_b^0 \to \Sigma_c^+\pi^0 l^-\bar{\nu}_l$, with
      $\Sigma_c^+ \to \Lambda_c^+ \pi^0$;
\item $\Lambda_b^0 \to \Lambda_c^{\ast +} l^- \bar{\nu}_l$, with 
      $\Lambda_c^{\ast +} \to \Lambda_c^+ \pi^+ \pi^-$, 
      or $\Lambda_c^+ \pi^0\pi^0$.
\end{itemize}
The search for resonant states described in this section is not 
sensitive to channels with $\pi^0$'s. 
The first three decay modes are expected to have the same branching 
fractions.
The $\Sigma_c$(2455) and $\Sigma_c$(2520) resonances would show up 
as peaks in the distribution of the variable $Q_{\Sigma}$, defined as:
\begin{eqnarray}
Q_{\Sigma} = M(\Lambda_{c}^+ \pi ) - M(\Lambda_{c}^+) -  m_{\pi}.
\end{eqnarray}
As the relative sign between the $\Lambda_c^+$ and the charged pion has 
not been distinguished, the $Q_{\Sigma}$ distribution should 
contain the same number of events from 
$\Lambda_b^0 \to \Sigma_c^{++}\pi^- l^- \bar{\nu}_l$, and
$\Lambda_b^0 \to \Sigma_c^0   \pi^+ l^- \bar{\nu}_l$, and
the same combinatorial background. For display purposes, 
the expected spectrum for the decay 
$\Lambda_b^0 \to (\Sigma_c(2455) \pi) l^- \bar{\nu}_l$
is shown as a dashed line in Figure~\ref{fig:3}, 
under the assumption that the observed number of events satisfies:
\begin{eqnarray}
\frac{N_{obs}(\Lambda_b^0 \to \Sigma_c^{++} \pi^- l^- \bar{\nu}_l) +
      N_{obs}(\Lambda_b^0 \to \Sigma_c^0 \pi^+ l^- \bar{\nu}_l)}
{N_{obs}(\Lambda_b^0 \to \Lambda_c^+ (X) l^- \bar{\nu}_l)} = 0.1
\end{eqnarray}
where the denominator is the total number of events in the 
$\Lambda_c^+ l^-$ sample.
The $\Sigma_c$(2520) signal, indicated as a dotted line 
histogram, is broadened due to its natural width of 20~MeV$/c^2$. 

%DB \begin{figure}[ht]
%DB
%DB \begin{figure}[ht]
\begin{figure}
\begin{center}
\epsfxsize = 9.5 true cm
%DB \epsfxsize = 8 true cm
\epsffile{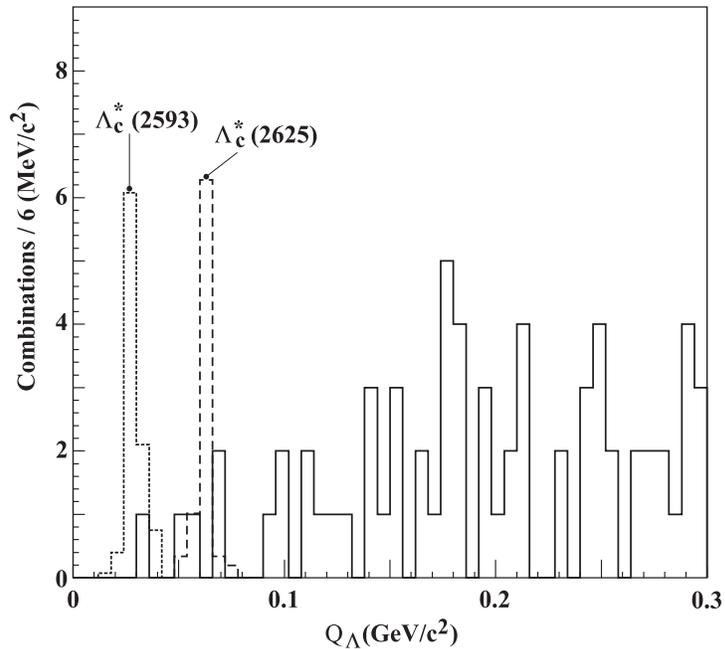}
\caption{\it 
Distribution of 
$Q_{\Lambda} = M(\Lambda_c^+ \pi\pi)-M(\Lambda_c^+)-2 m_{\pi}$ in data 
(solid line histogram). 
The simulated $\Lambda_c^{\ast}$ signals (dotted and dashed line histograms)
assume 
$N_{obs}(\Lambda_b^0 \to \Lambda_c^{*+} l^- \bar{\nu}_l) = 
0.1 \; N_{obs}(\Lambda_c^+ (X) l^- \bar{\nu}_l)$.}
\label{fig:4}
\end{center}
\end{figure}

From the number $N_{obs}$ of events found in the signal 
region $Q_{\Sigma}$ (column 3 of Table~\ref{tab:1}), 
upper limits have been derived on the number of observable decays 
from the $\Lambda_b^0 \to (\Sigma_c \pi)^+ l^- \bar{\nu}_l$ channels 
(Table~\ref{tab:1}).
These limits are obtained from the highest expected yield of signal 
events, $\bar{N}_{\Sigma}$, such that the probability: 
\begin{eqnarray}
Prob(N_{\Sigma}+ N_{bkg}\leq N_{obs}) > 5\%.
\end{eqnarray}

\begin{table}
\begin{center}
\begin{tabular}{|l|c|c|c|c|c|}    
\hline
Decay channel & $Q_B$ window& $N_{obs}$ & $N_{bkg}$ & $N_{limit}$ 
              & limit on \\
& (GeV/c$^2)$ & observed  & estimated & upper  & $R_{channel}$\\
\hline
$\Lambda_b^0 \to \Sigma_c(2455)^{++} \pi^{-} l^- \bar{\nu}_l \; + $ 
& 0.025-0.030 & 3 & 2.5 & 5.30 & 0.078\\
$\Lambda_b^0 \to \Sigma_c(2455)^0 \pi^+ l^- \bar{\nu}_l$ & & & & &\\
\hline
$\Lambda_b^0 \to \Sigma_c(2520)^{++} \pi^- l^- \bar{\nu}_l \; + $ 
& 0.080-0.105 & 15 & 10.5 & 12.5 & 0.190\\
$\Lambda_b^0 \to \Sigma_c(2520)^{0} \pi^+ l^- \bar{\nu}_l$ & & & & &\\
\hline
$\Lambda_b^0 \to \Lambda_c(2593)^{\ast +}  l^- \bar{\nu}_l$ 
& 0.024-0.036 & 1 & 0.5 & 4.30 & 0.064 \\
\hline
$\Lambda_b^0 \to \Lambda_c(2625)^{\ast +}  l^- \bar{\nu}_l$ 
&0.054-0.066 & 1 & 1.5 & 3.20 & 0.048\\
\hline
\end{tabular} 
\caption{\it $Q_B$ window ($B=\Sigma$, $\Lambda$), 
number of observed events $N_{obs}$, estimated 
number of background events $N_{bkg}$, 95\% C.L.
upper limit on the observable number of signal events
$N_{limit}$, and the upper limit on the channel contribution 
to the observed sample $R_{channel}$.}
\label{tab:1}
\end{center}
\end{table}

The probability law for the number of events is assumed to be a 
Poisson law, with the mean given by the sum of the background 
contribution $\bar{N}_{bkg}$, interpolated from the 
adjacent bins of the $Q_{\Sigma}$ spectrum, 
and $\bar{N}_{\Sigma}$, the expected yield. The uncertainty 
on the estimated mean background level has not been included.

To search for the $\Lambda_c^{\ast +}$ states, 
the distribution of the variable
\begin{eqnarray}
Q_{\Lambda} = M(\Lambda_{c}^+ \pi^+ \pi^-) - M(\Lambda_{c}^+) - 2 m_{\pi}
\end{eqnarray}
is considered (see Figure~\ref{fig:4}), 
and 95\% C.L. upper limits have been obtained using the same
procedure as in the $(\Sigma\pi)$  final states. 
The kinematical window for $Q_{\Lambda}$ is given by 
the column~2 of Table~\ref{tab:1}.\\
The upper limits in column~5 of Table~\ref{tab:1} refer to the observable 
final states. 
The contributing fraction $R_{channel}$ in column~6 is: 
\begin{eqnarray}
R_{channel} = \frac{N_{channel} }{N_{obs}(\Lambda_c^+ (X) l^- \bar{\nu}_l)}
\end{eqnarray}
where $N_{channel}$ is the (maximal) number of events from this 
channel contributing to the $\Lambda_c^+ l^-$ sample: a correction factor
of 3/2 has been applied to the upper limits for the 
$\Lambda_c^{\ast +}$ final states, as the pion pair is 
assumed to have $I=0$. The same factor 3/2 has been applied to the observed
$\Sigma_c^+$ contribution in order to include the $\Lambda_c^+ \pi^0$ 
final state.  

\subsection{ The ${\bold \Lambda_c^+ \pi \pi}$ contribution}

\begin{figure}[!h]
\begin{center}
%DB \epsfxsize = 16. true cm 
\epsfxsize = 17. true cm 
\epsffile{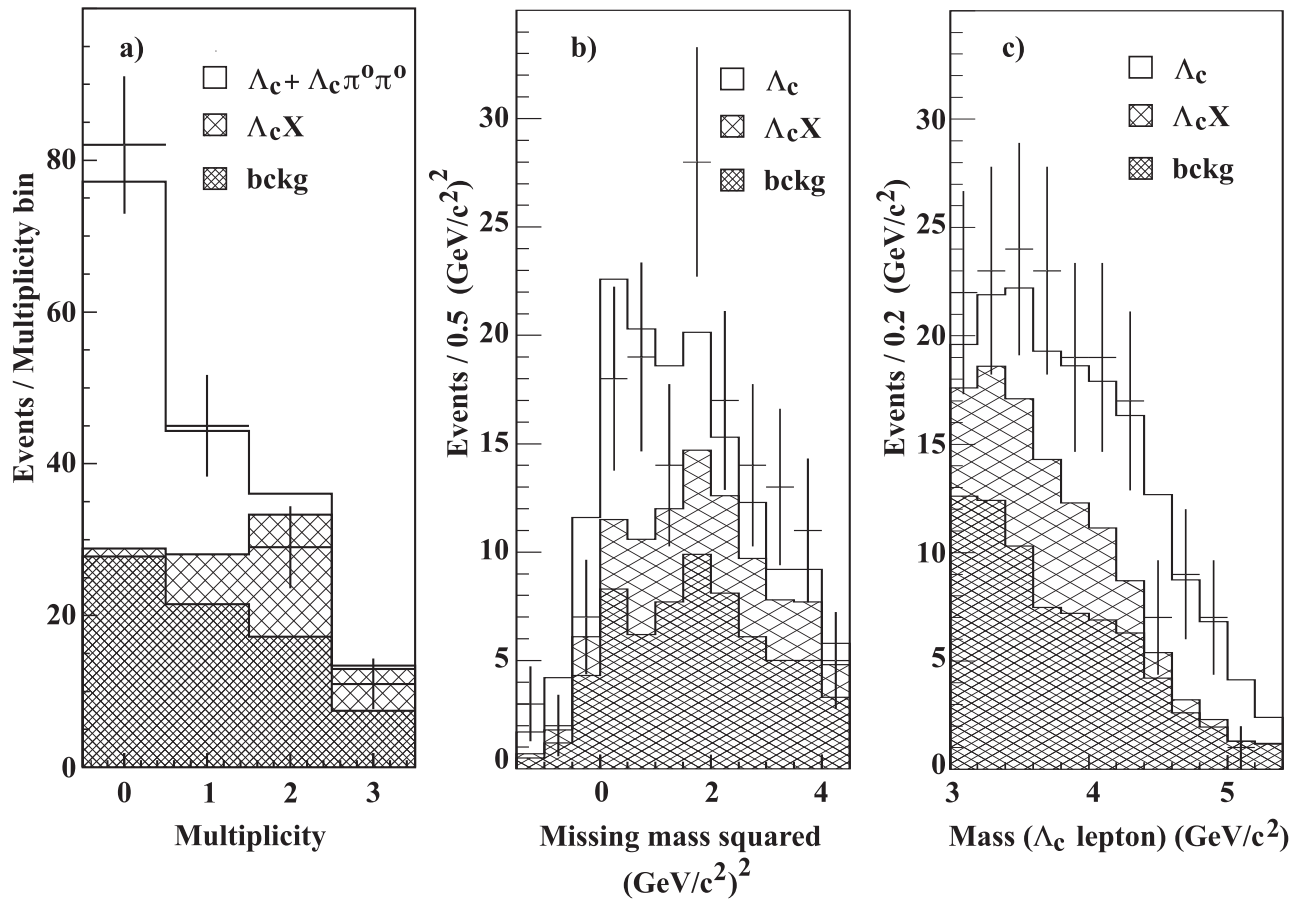}
\caption{ \it a) The charged multiplicity;
b) the missing mass squared and c) the ($\Lambda_c l$) mass distributions in the
$\Lambda_c^+ $ mass window. The hatched areas show the 
$\Lambda_c^+ \pi^+ \pi^-$ contributions (including $\pi^0 \pi^0$).
The cross hatched histogram is the combinatorial background.
}  
\label{fig:5}
\end{center}
\end{figure}

No evidence for the production of excited charmed states
in $\Lambda_b^0$ semileptonic decay has been found, but the limits obtained 
still allow for a substantial contribution from these final states. 
As, in addition, non-resonant charmed channels might be present,
an inclusive approach will be adopted to evaluate the combined 
contribution of the resonant and non-resonant final states. 
This inelastic contribution is investigated by studying  
three distributions:
%DB
\newpage
%DB
\begin{itemize}
\item the multiplicity $N_c$ of charged particle tracks associated 
to the secondary vertex (Figure~\ref{fig:5}a); 
\item the missing mass squared 
$M_{miss}^2 = (p_{\Lambda_b} - p_{\Lambda_c^+} - p_l)^2$
 (Figure~\ref{fig:5}b); 
\item the $\Lambda_c^+$-lepton mass, which is expected to be 
      smaller for $\Lambda_c^+ (\pi\pi) l^- \bar{\nu}_l$ 
final states (Figure~\ref{fig:5}c).
\end{itemize}

\subsubsection{The multiplicity} 

The multiplicity of charged particles other than the 
$\Lambda_c ^+$ decay products
and the lepton, compatible with the $\Lambda_b^0$ vertex was
evaluated by a neural network algorithm which separates tracks 
compatible with the primary or the secondary vertex. 
The probability of the vertex assignment was required to be larger 
than 0.5.  The $b$ and $c$ vertices were not separated in this treatment.
The result for the events situated in the $\Lambda_c^+$ mass window 
 2.260 GeV/$c^2 < M_{\Lambda_c} < 2.310$ GeV/$c^2$ 
is shown in Figure~\ref{fig:5}a. The contribution of the  
$\Lambda_c^+ \pi^0 \pi^0 l^- \bar{\nu}_l$ channel is assumed to be 
1/2 of the charged pion mode $\Lambda_c^+ \pi^+ \pi^- l^- \bar{\nu}_l$,
as the hadronic state is assumed to have $I = 0$. 

\subsubsection{The missing mass in 
${\bold \Lambda_b^0 \to \Lambda_c^+ l^- \bar{\nu}_l}$}

To reconstruct the missing mass squared $M^2_{miss}$, with 
$M^2_{miss} = (p_{\Lambda_b} - p_{\Lambda_c} - p_l)^2$, 
the four-momentum of the $\Lambda_b^0$ was evaluated assuming 
the decay channel $\Lambda_b^0 \to \Lambda_c^+ l^- \bar{\nu}_l$.
The energy is  
$ E_{\Lambda_b} = E_{\Lambda_c} + E_l + E_{\nu}$,
and the $\Lambda_b^0$ momentum has been computed as 
$\sqrt{E_{\Lambda_b}^2 -m_{\Lambda_b}^2}$
with $m_{\Lambda_b} = 5.624$ GeV$/c^2$.  
The neutrino energy and the direction of the $\Lambda_b^0$ have
to be determined. The energy of the undetected neutrino in the 
exclusive $\Lambda_b^0 \to \Lambda_c^+ l^- \bar{\nu}_l$ channel was 
estimated as in previous studies of semileptonic decay 
of $B$ mesons~\cite{ref:delphivcb1}. 
The total energy in the $\Lambda_c^+ l^-$ hemisphere, which is the sum 
of the visible energy ($E_{same}$) in the same hemisphere,
and of the neutrino energy ($E_{miss}$), was computed 
using the observed total masses of the $\Lambda_c^+$ hemisphere ($M_{same}$)
and of the opposite hemisphere ($M_{oppo}$), together with four-momentum 
conservation:  
\begin{eqnarray}
E_{miss} + E_{same} = \frac{\sqrt{s}}{2} + 
{{M_{same}^2-M_{oppo}^2}\over{2\sqrt{s}}}.
\end{eqnarray}

As $E_{same}$, $M_{same}$, and $M_{oppo}$, are approximately known
due to the detector inefficiencies, an empirical correction 
$f_{sim}(E_{same}$) (as in~\cite{ref:delphivcb1}) has been estimated from
a simulation of the exclusive semileptonic channel, to 
improve the accuracy on the neutrino energy reconstruction:
\begin{eqnarray}
E_{\bar{\nu}_l} = E_{miss} + f_{sim}(E_{same}).
\end{eqnarray}
Whenever this procedure leads to a negative energy, the value  
$E_{\bar{\nu_l}} = 0$ is used. The final resolution  
on the neutrino energy is around $33\%$. 
Adding this energy to the energy of the $\Lambda_c^+$ and of 
the lepton gives the energy of the $\Lambda_b^0$ with a resolution of 
8$\%$ (Gaussian fit).\\
\noindent   
The direction of the $\Lambda_b^0$ can be estimated by two different 
methods: the momentum of the $\Lambda_c^+ l^-$ system
gives the best accuracy when the line of flight is short, 
while the positions of the primary and secondary vertices,
work best at large separations. In the data from 1992 and 1993, 
where the microvertex detector did not provide any $z$ information, 
the $\Lambda_c^+ l^-$ momentum was always used to measure 
the polar angle $\theta$. The weighted combination of the two 
which gave the best resolution on the Isgur-Wise variable $w$
was chosen. 

It is seen in Figure~\ref{fig:5}b that 
the reconstructed missing mass squared $M^2_{miss}$ is sensitive to the 
presence of the inelastic channel.

\subsubsection{The ${\bold \Lambda_c^+ l}$ mass}

The $\Lambda_c^+l^-$ mass distribution 
is expected to be shifted to lower values in 
the inelastic $\Lambda_c^+ \pi\pi l^- \bar{\nu}_l$ final states. 
This effect is apparent in Figure~\ref{fig:5}c and it constrains the 
proportion of inelastic events

\subsubsection{Fit of the elastic fraction ${\bold f_{\Lambda_c}}$} 

To measure the fraction $f_{\Lambda_c}$ of elastic 
$\Lambda_c^+ l^- \bar{\nu}_l$ decays, the three previous distributions
are considered to be the sum of three components:
\begin{itemize}
\item the $\Lambda_b^0 \to \Lambda_c^+ l^- \bar{\nu}_l$ decays;
\item the $\Lambda_b^0 \to \Lambda_c^+ \pi\pi l^- \bar{\nu}_l$ decays. 
 As this channel is a sum of many states, it is simulated without a form 
 factor, using the quark matrix elements for weak decays in PYTHIA
\cite{ref:sjostrand2}; 
\item events from the combinatorial background present under the 
 $\Lambda_c^+ $ signal. 
 The shape of this component is evaluated using events 
 situated in the side bands of the $\Lambda_c$ mass peak.
 Its normalisation is fixed according to the fit of Figure~\ref{fig:1}.
\end{itemize} 
The overall normalisation is fixed to the total number of observed events.   
The elastic fraction   $f_{\Lambda_c}$  
in the final state is obtained from three fits to the three 
distributions in Figures~\ref{fig:5}. Each of them is
first adjusted independently. The statistical correlations 
of the three results are  obtained from the simulated two 
dimensional distributions of the three pairs of variables.
They are respectively:
0.25 for (multiplicity, $\Lambda_c^+ l^-$ mass), 0.35 for
(multiplicity, missing mass), and 0.65 for (missing mass, 
$\Lambda_c^+ l^-$ mass). The combined fit to the three values gives:
\begin{eqnarray}
f_{\Lambda_c} = \frac{N(\Lambda_c^+ l^- \bar{\nu}_l)}
{N(\Lambda_c^+ l^- \bar{\nu}_l) + N(\Lambda_c^+ \pi\pi l^-\bar{\nu}_l)} =
0.60 \pm 0.08 \;(\hbox{stat}) \pm 0.06 \;(\hbox{syst}).
\end{eqnarray}
The ratio $f_{\Lambda_c}$ is {\it not} a branching fraction, but 
the ratio of the observed contributions in the data sample. 
The systematic uncertainties arising 
from the identification efficiency, the time of flight and the 
modelling of the $b$-quark fragmentation are negligible. 
The uncertainty on $\ro$, as measured in this paper, 
changes $f_{\Lambda_c}$ by $\pm 0.02$, and the uncertainty from the
simulation of the inelastic channels is estimated 
to be $\pm 0.03$ by substituting a $\Sigma_c \pi $
decay for  the $\Lambda_c^+ \pi\pi$ non-resonant prediction. 
The systematic uncertainty from the combinatorial background is estimated 
by comparing side bands of different sizes and contributes $\pm 0.05$. 
The limits on the observed resonance contributions given in section~6.1 
can be turned into limits on branching fractions.  
The fraction of observable decays can be derived
from the assumption of an $I=0$ final state and is found, as already 
mentioned, to be 2/3. The kinematical acceptances for 
$(\Sigma_c\pi,\Lambda_c^{\ast}) l^- \bar{\nu}_l$
and $\Lambda_c^+ l^- \bar{\nu}_l$ in the charged 
decay channels differ, with 
$\epsilon(\Lambda_c^+ (\pi^+\pi^-)l^- \bar{\nu}_l)
/\epsilon(\Lambda_c^+ l^- \bar{\nu}_l) = 0.57$,
according to the simulation. 
The branching fraction to the 
exclusive $\Lambda_c^+ l^- \bar{\nu}_l$ final state can  
then be estimated to be:
\begin{eqnarray}
\frac{\Gamma(\Lambda_b^0 \to \Lambda_c^+ l^- \bar{\nu}_l)}
{\Gamma (\Lambda_b^0 \to \Lambda_c^+ l^- \bar{\nu}_l) + 
\Gamma(\Lambda_b^0 \to \Lambda_c^+ \pi\pi l^- \bar{\nu}_l)} 
= 0.47 ^{+0.10}_{-0.08} \;(\hbox{stat})  ^{+0.07}_{-0.06} \;(\hbox{syst}).
\end{eqnarray}
Although half of the $\Lambda_c^+l^-$ events in the $\Lambda_b^0$ decay 
arise from a $\Lambda_c^+ \pi \pi l^- \bar{\nu}_l$ final state,
no signal from resonant production of $\Sigma_c$ or $\Lambda_c^{\ast}$
has been observed in the present sample. 

\section {Leptonic branching fraction from the event rate}

The previous determination of the fraction of elastic semileptonic
decays of $\Lambda_b^0 \to \Lambda_c^+ l^- \bar{\nu}_l$ in the
$\Lambda_c^+ l^-$ sample allows a measurement of the semileptonic 
branching fraction, $B_{\Lambda_c}$, in this transition. 
As the total width of the $\Lambda_b^0$ is given by its lifetime, 
this branching fraction can then provide a direct measurement of 
the slope parameter $\ro$ of the form factor. 
The branching fraction $B_{\Lambda_c}$ has been  
measured from the number 
of $\Lambda_b^0 \to \Lambda_c^+ l^- \bar{\nu}_l$ candidates
using only the decay channel $\Lambda_c^+ \to p K^- \pi^+$, 
which has less background: 
\begin{eqnarray}
N (\Lambda_c^+ l^- \bar{\nu}_l) = N_Z^h \; 2 \; R_b 
\; f(b \to \Lambda_b^0)
\; B_{\Lambda_c} \; Br (\Lambda_c^+ \to p K^- \pi^+)
\; \epsilon (p K \pi l) \; N_l.
\end{eqnarray}
As the event rate is much more sensitive than the $w$ shape to 
the knowledge of the identification efficiencies, this analysis has  
been restricted  to the periods of data taking where
both gas and liquid RICH were present. The  number of  hadronic 
$Z$ decays is $N_Z^h = 1.52 \times 10^6$, and the fraction of $b$-flavoured 
final states is $R_b = 0.217$~\cite{ref:pdg}.
Detailed comparisons have shown that the identification efficiencies 
in data and simulation are then in excellent agreement. 
The simulated reconstruction and identification efficiency 
is found to be, including the lepton,
$\epsilon(p K \pi l) = (8.0 \pm 0.8)\% $. 
The branching fraction  
$Br (\Lambda_c^+ \to p K^- \pi^+) = (5.0 \pm 1.3)\%$, and 
$f(b \to \Lambda_b^0) = 0.108 \pm 0.020$ (subtracting 1\% 
for $\Xi_b$ from the quoted value in~\cite{ref:pdg}) are used.
The number of lepton families is $N_l = 2$,
as the $\tau$ contribution is negligible. 
The total number of observed $\Lambda_c^+ l^- \bar{\nu}_l$ 
and $\Lambda_c^+ X l^- \bar{\nu}_l$ 
events (without the enrichment selections for the elastic channel
of section~8)
is $ 47 \pm 10$. 
The fake lepton contamination, as determined in section~4.2 
amounts to $2.0\pm0.7$ events and should be subtracted. 
The ``observed" number $N(\Lambda_c^+ l^- \bar{\nu}_l)$ 
of exclusive $\Lambda_c^+ l^- \bar{\nu}_l$ decays can be 
estimated from the latter fraction to be
$N_{obs} = f_{\Lambda_c} \cdot (47 -2.0) = 27 \pm 6$. 
This implies $B_{\Lambda_c} = (4.7 \pm 1.1\;\hbox{(stat}))\%$. 
The main sources of systematic errors  on $B_{\Lambda_c}$ 
are given in Table~\ref{tab:2} and sum to a relative error of 
37\%, so that:
\begin{eqnarray}
B_{\Lambda_c} = (4.7 \pm 1.1 \;\hbox{(stat)} \pm 1.7 \;\hbox{(syst)})\%.
\end{eqnarray}
Most of the systematic error arises from the uncertainty on
$Br(\Lambda_c^+ \to p K^- \pi^+)$. The other systematic errors include
the uncertainty on the reconstruction efficiency $\epsilon$, including
identification, the contribution of 
the $B$-meson decays which was varied up to the maximal value of 
3.4\% found in section~5. The impact of the lifetime uncertainty on the 
acceptance is negligible, but its effect on the parameter $\ro$ via the
normalisation of the branching fraction is large and it is included
here to simplify the presentation. 
The contribution of the $ pK^- \pi^+$  branching ratio can be explicitly 
extracted:
\begin{eqnarray}
B_{\Lambda_c} = (4.7 \pm 1.1 \;\hbox{(stat)}
\pm 1.3 \;(Br(\Lambda_c \to p K^- \pi^+))
\pm 1.3 \;\hbox{(other syst))\%}.
\end{eqnarray}
 
The semileptonic decay width can be computed from  $\ro$, using equation~(2),
under the assumption that $\xi_B(1) = 1$. The total width
is given by the $\Lambda_b^0$ lifetime 
$\tau_{\Lambda_b} = 1.23 \pm 0.08$~ps~\cite{ref:pdg},
and the semileptonic 
branching fraction $B_{\Lambda_c}$ provides (within HQET) an estimate 
of the slope $\ro$:
\begin{eqnarray}
\ro_{rate} = 2.05^{+0.70}_{-0.50} \;\hbox{(stat error only)}.
\end{eqnarray}

%DB \newpage
\begin{table}
\begin{center}
\begin{tabular}{|l|c|c|c|}   
\hline
reference value & $\pm$ uncertainty range & $ \delta B_{\Lambda_c}/B_{\Lambda_c}$\\
\hline
$Br(\Lambda_c^+ \to p K^- \pi^+) = 0.050$ & $  0.013 $     &  0.26 \\
$f(b \to \Lambda_b^0) = 0.108$             & $ 0.020 $     &  0.18 \\
$f_{\Lambda_c} = 0.60 $                    & $  0.10$       & 0.16 \\
$\epsilon(p K \pi l)= 0.080$               & $ 0.008$      &  0.11 \\
$\tau_{\Lambda_b} = 1.23 $~ps            & $  0.08$~ps   & 0.065 \\
$B$ meson decays $< 3.4 \%$              & $  0.034 $     & 0.034 \\
fake leptons = 4.3\%                      & $ 1.5\%$     & 0.016 \\ 
\hline 
{\bf Total } ($\delta B_{\Lambda_c}/B_{\Lambda_c}$)  &  & 0.37\\
\hline
\end{tabular} 
\caption{\it Main sources of systematic uncertainties on the expected rate.}
\label{tab:2}
\end{center}
\end{table}

This measurement of $\ro$ will be combined in the next section 
with a fit to the distribution of the Isgur-Wise variable $w$
to obtain an improved determination of the slope parameter and 
of the branching fraction. The impact of systematic errors will 
be evaluated for this combined fit. 

\section{Combined fit to the ${\bold w}$ shape and the event rate}

\subsection {The enriched ${\bold \Lambda_c^+}$-lepton sample}

As has been shown in Figure~\ref{fig:1}, the number of   
$\Lambda_c^+$-lepton pairs is $101 \pm 17$, and this 
sample contains the elastic $\Lambda_c^+ l^- \bar{\nu}_l$ 
and inelastic $\Lambda_c^+ (\pi\pi) l^- \bar{\nu}_l$ final states.
The charged multiplicity $N_c$, as well as the 
($\Lambda_c^+$ lepton) mass can be used to enrich the 
sample with respect to the elastic channel, by selecting
multiplicities smaller than 2, and 
$\Lambda_c^+ l^-$ masses larger than 3.5~GeV/$c^2$. 
The number of $\Lambda_c^+$ candidates left is
obtained from a fit to the mass distribution of the candidates,
shown in Figure~\ref{fig:6}, and is 
$62.5 \pm 10.0$. The remaining $\Lambda_c \pi\pi$ contribution
is obtained from the simulated efficiency, after application 
of the enrichment selection, and is found to be $10.0 \pm 2.8$ events.  
The fraction of elastic $\Lambda_c^+ l^- \bar{\nu}_l$ events in
this sample is $r_{\Lambda_c} = 0.84 \pm 0.17$.
The $\Lambda_c$ mass distribution of the enriched sample,
together with the different components is shown in Figure~\ref{fig:6}. 
The slope parameter of the $w$ distribution is
determined from this enriched sample. 
The candidates obtained from the three decay modes of the 
$\Lambda_c^+$ considered in section~4 
have been separately analysed. In each of these samples,   
the non-$\Lambda_c^+$ background is measured from a fit to the 
$\Lambda_c^+$ mass distribution.

The complementary sample, enriched in $\Lambda_c^+ \pi\pi$, will be  
used to monitor the $w$-shape of the $\Lambda_c^+ \pi\pi$ background. 

\subsection{The ${\bold w}$-shape likelihood}

The four-momentum of the $\Lambda_b^0$ meson is reconstructed
as described in section~6.2.2.  
The values of $q^2$ and $w$ can then be estimated. 
The resolutions achieved are similar to those obtained 
in~\cite{ref:delphivcb2}, 
and $\Delta w/w$ is close to $\pm 8\%$. 

A likelihood fit to the two-dimensional $(M({\Lambda_c}),w)$ distribution 
is then performed, with $2.190 < M({\Lambda_c}) < 2.385$~GeV/$c^2$ 
and $1.0 < w < 1.6$. 
The mass and $w$ dependences are assumed to factorise 
in the probability distribution for each event 
$k$ in channel $i$,
where the index $i = 1,2,3$ runs over the three input channels  
(three final states, the two lepton samples are combined): 
\begin{eqnarray}
P_i(M_k,w_k) = f^S_i\,S_i(w_k,\ro)G(M_k) +
f^{\Lambda_c \pi\pi}_i\,B_i^{\Lambda_c}(w_k)G(M_k)
+ f^B_i\, B_i^{no\Lambda_c}(w_k).
\end{eqnarray}

\begin{figure}[h]
\begin{center}
%DB \epsfxsize = 10 true cm
\epsfxsize = 12 true cm
%DB \epsfxsize = 8 true cm
\epsffile{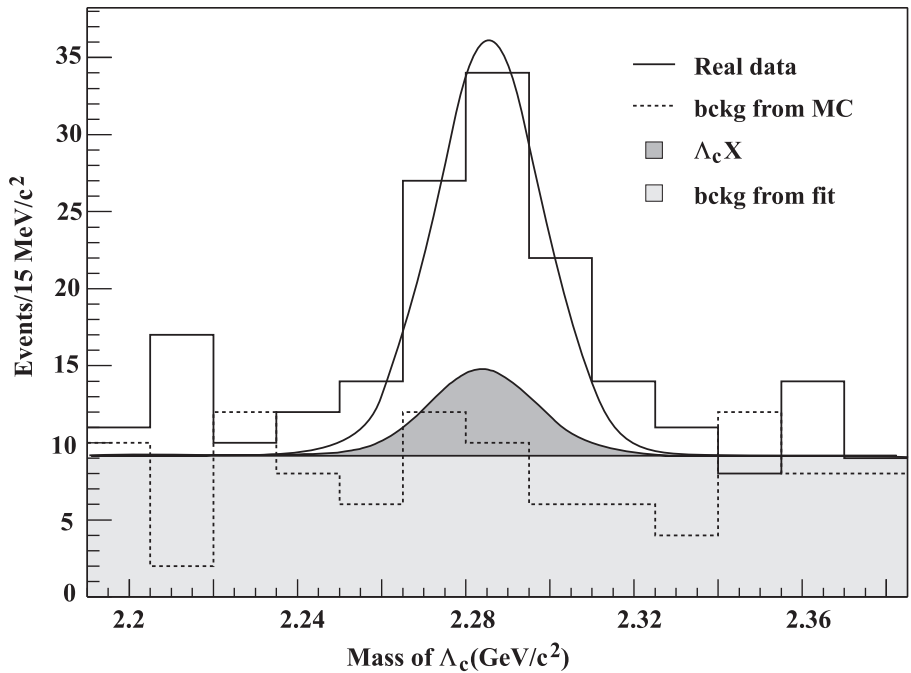}
\caption{\it 
The $\Lambda_c$ mass distribution  
in the (enriched/elastic) sample of $\Lambda_c^+l^-$ events
used in the likelihood shape analysis (all channels). 
The estimated background
(dotted histogram) 
%DB 
is obtained from simulated hadronic $Z$ decays and normalised to
the actual data}
\label{fig:6}
\end{center}
\end{figure}

The Gaussian term $G(M_k)$ describes the $\Lambda_c^+$ contribution,
$B_i^{\Lambda_c}(w)$ is the inelastic background,
while the combinatorial background $B_i^{no\Lambda_c}(w)$ 
is mass independent in the $\Lambda_c$ mass window 
investigated, as justified by a direct inspection of its
shape in the simulation in Figure~\ref{fig:6}.  
The coefficients $f_i^s$ (for the elastic 
 $\Lambda_c$ signal), $f_i^{\Lambda_c \pi\pi}$, and  
$f_i^B$ are fixed, and are obtained from the $p K \pi$ mass spectrum
and the elastic fraction of 0.84. 
 
In each channel $i$, the contribution of the signal $S_i(w)$ is 
obtained from the simulated events weighted at a given $\ro$ by 
the squared form factor:
$\exp(-2 \ro (w_g - 1)) $ where $w_g$ is the generated
value of $w$. 
The function $S_i(w,\ro)$ 
is a convolution of the physical $w$ distribution with the 
detection efficiency and the resolution of the reconstruction
of this variable. It has been expressed as a factorised expression,
as it was easier to parametrise directly 
$S_i$ rather than the resolution function:
\begin{eqnarray}
S_i(w ,\ro ) = S_i(w,\ro = 0)\; \exp(\ro \cdot\alpha(w)). 
\end{eqnarray}
The function $S_i(w,\ro = 0)$ is parametrised as 
$S_i(w, \ro = 0) = a (w-1)^m \exp(p(w-1) + q(w-1)^2))$, with 
$a = 0.458$, $m = 0.154$, $p = 17.2$, $q = 40.3$. 
A very good description of the simulated $w$ distributions 
was obtained when the exponent $\alpha(w)$ was assumed to be a 
linear function of $w$: 
\begin{eqnarray}
\alpha(w) = \alpha_0 + \alpha_1 \cdot (w-1). 
\end{eqnarray}
The coefficient $\alpha_0$ is a normalisation coefficient, while
$\alpha_1$ describes the $w$ dependence of the signal. In the  
absence of smearing and detector effects, $\alpha_1$  would be -2,
given the exponential parametrisation in equation~(7), and
the actual value $\alpha_1 = -1.67 \pm 0.08$ is close to this 
estimate.  

The $B_i^{\Lambda_c}(w)$ and $B_i^{no \Lambda_c}(w)$ functions 
are found from the data. The shape of $B_i^{\Lambda_c}(w)$ is 
obtained from the $w$ distribution of the subsample of 
$\Lambda_c^+$ events enriched in $\Lambda_c^+ \pi\pi$, with
a charged multiplicity at the $\Lambda_b^0$ vertex $N_c \geq 2$ 
or with a $(\Lambda_c^+ l^-)$ mass $< 3.5$~GeV$/c^2$.
A correction factor derived from the simulation is applied to the
$w$ distribution of this sample, to relate the background shapes
in the enriched (elastic) and anti-enriched ($\Lambda_c \pi\pi$) 
samples. This background, corrected for the enrichment bias is 
then described by a function 
$B^{\Lambda_{c}}(w) = (w-1)^{a_1} \exp(-b_1(w - 1))$,
with $a_1 = 2.22 \pm 0.50$, 
and $b_1 = 27.5 \pm 5.8$. \\
The non-$\Lambda_c^+$ background, $B_i^{no\Lambda_c}(w)$,
is evaluated from side-bands 
in the mass spectrum, chosen outside the mass window 
$2.260 < M(\Lambda_c) < 2.310 \;\hbox{GeV}/c^2$.
Its shape is described by the same parametrisation
as the previous one, with new values of $a$ and $b$. 
In the dominant $p K^- \pi^+$ channel, for instance:
$a = 4.28 \pm 0.34$, and $b = 22.3  \pm 1.9$.

The distributions of the signal in the different channels $S_i(w,\ro)$,
and the shapes of the backgrounds $B_j(w)$ are 
normalised to unity, and the coefficients $f^S_i$, $f^B_j$ have been 
measured from the data and are fixed. 
The likelihood is the sum: \\
${\cal L}_{shape} = -\sum_{k} \mbox{Log}(P(M_k,w_k))$, and 
the slope obtained from the one parameter fit to the $w$ distribution is:
\begin{eqnarray}
\ro = 1.59 \pm 1.10 \;\hbox{(stat)}. 
\end{eqnarray}
The quality of this fit can be checked in Figure~\ref{fig:7}a, where the 
predicted distribution has been normalised to the observed 
number of events in real data. The $\chi^2/NDF$ between the 
distribution predicted from the likelihood fit
and the observed distributions is $5.2/11$. 

\subsection{The combined event rate and ${\bold w}$ shape likelihood}

The information on the shape and on the absolute rate can be combined 
into an optimised determination of $\ro$, assuming $\xi_B(1) = 1$, 
and a new likelihood fit is performed where the observed number of 
events is included as a constraint. 
The expected number of events $N_{ex}$ is derived from 
the semileptonic branching fraction, which is itself a function of 
$\ro$ as described in section~7:  
\begin{eqnarray}
{\cal L} = {\cal L}_{shape} - N_{obs}\mbox{Log}(N_{ex}) + N_{ex}.
\end{eqnarray}

\begin{figure}[h]
\begin{center}
\epsfxsize=16.cm 
%DB \epsfxsize=16.cm 
\epsffile{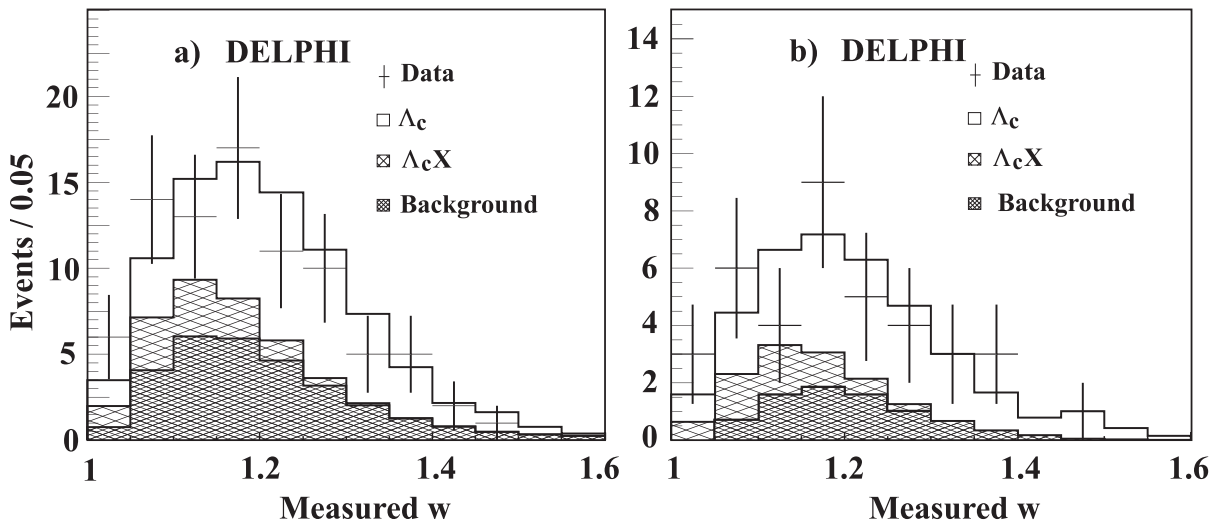}
\caption{\it Comparison between fitted and measured 
$w$ distributions inside the $\Lambda_c$ mass window:
a) fit to the $w$ distribution;
b) combined fit to the $w$ distribution and 
the $\Lambda_c^+ l^-\bar{\nu}_l $ rate (restricted to the working
RICH sample). The statistics in Figure~\ref{fig:7}b are smaller 
than in \ref{fig:7}a.}
\label{fig:7}
\end{center}
\end{figure}

The full statistics are used for the shape likelihood, while
as in section~7, the rate $N_{ex}$ is measured
only when the gas and liquid RICH are simultaneously  operational,
and for the $\Lambda_c^+ \to p K^- \pi^+$ channel alone
~\footnote{The statistics of observed and expected events were scaled 
to reproduce the actual statistical uncertainty on the number
of events.}.
The actual likelihood function includes the contributions
from the different backgrounds, and the value of $\ro$ obtained 
from this one parameter fit, shown in Figure~\ref{fig:7}b, is:
\begin{eqnarray}
\ro = 2.03 \pm 0.46 \;\hbox{(stat)}.
\end{eqnarray}
The statistics in Figure~\ref{fig:7}b are smaller 
than in~\ref{fig:7}a as only data
with an operational RICH are used. 

The semileptonic branching fraction corresponding to this value of
$\ro$ is: 
\begin{eqnarray}
Br(\Lambda_b^0  \to \Lambda_c^+ l^- \bar{\nu}_l) = 
(5.0^{+1.1}_{-0.8} \;\hbox{(stat)})\%. 
\end{eqnarray}
The expected number of events is  $27^{+7}_{-4}$, while the observed  
number of elastic $\Lambda_c$ events quoted in section~7,
and included in the likelihood fit, is $27 \pm 6$.  

There are three groups of systematic errors: the errors associated to
the prediction of the $w$ shape
(the identification efficiency, 
the parametrisation of the shape as a function of $\ro$, 
the fragmentation), the uncertainties related to the expected 
yield (absolute efficiencies and branching fractions) 
and the systematic effects arising  from the subtraction of the 
$\Lambda_c$ background processes discussed in section~8.2
(parameters $a,a_1$ and $b,b_1$).
\par
 The momentum dependent uncertainties are 
dominated by the contribution to the global identification efficiency 
of the proton and kaon identification efficiencies
$\epsilon_{id}$. The observed momentum distributions 
and therefore the $w$ spectrum are sensitive to the detailed 
simulation of the identification algorithms.
The difference between the efficiencies of proton identification 
in data and in simulation was monitored using   
a sample of selected $\Lambda$ events.
For each year, the identification
efficiency as a function of momentum was measured for data and simulation.
The difference typically reaches 15\%. Simulated events were 
reweighted by the ratio of the data and simulation efficiency,
and the systematic error on $\ro$ 
measured from the shift in $\ro$ between the original and 
reweighted $w$ distributions.  

The resolution on $w$ is found to change by approximately 10\%
between different versions of the energy reconstruction algorithms.
The impact of this uncertainty was evaluated by degrading 
the resolution by 10\% in the simulation.
Given the small contribution of the $\Lambda_b^0$ semileptonic decays 
with a $\tau$  lepton, no systematic 
error was assigned for this component of the signal.

\begin{table}
\begin{center}
\begin{tabular}{|l|c|c|}   
\hline
source of error & $\pm$ uncertainty range & contribution to $\delta \ro$ \\
\hline
$\delta \epsilon_{id}/\epsilon_{id}$ & 0.15 & 0.08\\
$\delta w/w = 0.08  $  & 0.008 & 0.05 \\
$\alpha_1 = -1.67 $ & $0.08$ & 0.02 \\
Fragmentation ($<x_{\Lambda_b}>$) & $0.008$  & $0.05$ \\
$a,a_1$  and $b,b_1$ (shape)& stat. errors & 0.01 \\
Enriched fraction $r_{\Lambda_c}= 0.84$    & 0.17 & 0.10\\
Anti-enriched background shape & stat errors & 0.02 \\
\hline
$\delta B_{\Lambda_c}/B_{\Lambda_c}$ (from Table 2)
& 0.37   & $^{+0.70}_{-1.00}$ \\
\hline 
 {\bf Total} &  & $^{+0.72}_{-1.00}$\\
\hline
\end{tabular} 
\caption{\it Main sources of systematic uncertainties 
on the slope $\ro$.}
\label{tab:3}
\end{center}
\end{table}

The uncertainty on the number of $\Sigma_c$ and $\Lambda_c^{\ast}$ 
final states is included in the 
uncertainty on the fraction $r_{\Lambda_c}$, equal to the 
statistical uncertainty on its value. 
The systematic uncertainty on  the shape of the non-$\Lambda_c$ 
background is evaluated by changing the shape of this background 
according to the statistical error on its parametrisation. 
This background is evaluated from the side-bands 
of the $\Lambda_c$ mass spectrum. 

The main systematic uncertainties on $\ro$ are summarised in 
Table~\ref{tab:3}, where the dominant contribution of the normalisation 
error (as estimated in Table 2) is singled out. 
The value of $\ro$ is found to be:
\begin{eqnarray}
\ro = 2.03 \pm 0.46 \; \hbox{(stat)} \;\; ^{+0.50}_{-0.60}\;
(Br(\Lambda_c \to p K^- \pi^+)) \;\; ^{+0.50}_{-0.80}\; 
\hbox{(other syst)}.
\end{eqnarray}
The asymmetry in the errors on $\ro$ arises from the strong non-linearity
of the relation between the normalisation and the slope over the 
large range of $\delta B_{\Lambda_c}$. The contribution of 
the branching fraction into $p K \pi$ to 
the uncertainty on the normalisation is
given in Table 2, and the associated systematic error
on $\ro$ has been evaluated separately in equation (28).
The corresponding elastic semileptonic branching fraction is:
\begin{eqnarray}
Br(\Lambda_b^0  \to \Lambda_c^+ l^- \bar{\nu}_l)
 = (5.0^{+ 1.1}_{-0.8} \;\hbox{(stat)}^{+1.6}_{-1.2}\;\hbox{(syst)})\%.
\end{eqnarray}
To obtain $\ro$, the prediction of HQET $\xi(1) = 1$ 
has been assumed, and no theoretical 
uncertainty has been included in the systematic error to account
for this hypothesis. A 10\% change of $\xi(1)$ would amount
to a 20\% change in the rate, and to 0.3 in $\ro$.  

\section{Conclusions}

A first measurement of the form factor of the $\Lambda_b^0$ 
beauty baryon has been achieved 
in the $\Lambda_b^0 \to \Lambda_c^+ \l^- \overline{\nu_l}$ decay channel.  
Assuming an exponential behaviour of the Isgur-Wise function: 
\begin{eqnarray*}
\xi_B(w) = \xi_B(1 ) e^{-\ro(w - 1)},
\end{eqnarray*}
\noindent
the slope parameter $\ro$ was determined from a $w$-shape analysis to be:
$\ro = 1.59 \pm 1.10 \;\hbox{(stat)}$. If the validity of the HQET relation 
$\xi(1) = 1 $ is assumed, and the event rate taken into account, 
an improved determination of the slope can be obtained: 
\begin{eqnarray*}
\ro = 2.03 \pm 0.46 \;\hbox{(stat)}\;  ^{+0.50}_{-0.60} \;
(Br(\Lambda_c \to p K^- \pi^+))\;  ^{+0.50}_{-0.80} \;
\hbox{(other syst)}.
\end{eqnarray*}
The evaluation of the systematic errors takes into account the
actual variation of $\ro$ arising from each source. 

The semileptonic branching fraction into the exclusive semileptonic
 mode was measured within the hypothesis $\xi_B(1) = 1$ to be:
\begin{eqnarray*}
Br(\Lambda_b^0  \to \Lambda_c^+ l^- \bar{\nu}_l)
 = (5.0^{+1.1}_{-0.8} \;\mbox{(stat)} ^{+1.6}_{-1.2} \;\mbox{(syst)})\%.
\end{eqnarray*}

The fraction of elastic $\Lambda_c^+ l^-\bar{\nu}_l$ events is found to be:
\begin{eqnarray*}
\frac{\Gamma(\Lambda_b^0 \to \Lambda_c^+ l^- \bar{\nu}_l)}
 {\Gamma (\Lambda_b^0 \to \Lambda_c^+ l^- \bar{\nu}_l)+
 \Gamma (\Lambda_b^0 \to \Lambda_c^+ \pi\pi l^- \bar{\nu}_l)}
  = 0.47^{+ 0.10}_{-0.08} \;\mbox{(stat)} ^{+ 0.07}_{-0.06} 
\;\mbox{(syst)}\; . \end{eqnarray*}

The spectra of multiplicity, missing mass, and $\Lambda_c^+ l^-$ mass
shown in this paper strongly $\;\;$ hint at such a sizeable fraction of
non-elastic hadronic modes. This inelastic contribution is larger than 
assumed in the experimental determinations of the $\Lambda_b$ lifetime,
such as \cite{ref:delphilifetime}, and will affect its value, as the
ratio of the $\Lambda_c$ and $\Lambda_b$ momenta 
$p_{\Lambda_c}/p_{\Lambda_b}$ is 20\% lower in the inelastic
channel.  

In spite of the evidence for a large inelastic contribution
 in semileptonic decays, no indication of charmed baryon resonances 
 has been found in the final state.

The parameter $\ro$ reflects the structure of the $\Lambda_b^0$ 
baryon. Its value is somewhat larger than in the $b$-meson channel, where
$\ro \sim (0.60\; \hbox{to}\; 1.3)$ 
as measured in~\cite{ref:alephvcb,ref:opalvcb,ref:delphivcb2}.
A recent result on the $B$ meson decays from~\cite{ref:cleovcb}
suggests an even higher value of the slope of the Isgur-Wise 
function for mesons, 
with $\ro = 1.61 \pm 0.09 \;\hbox{(stat)} \pm 0.21 \;\hbox{(syst)}$. 
In all models proposed   
so far~\cite{ref:chakraverty,ref:sadzikowski,ref:cardarelli,ref:models}, 
the value of $\ro$ is 
expected to be larger for baryons.

%DB
\newpage
%         Modified on 04-06-1999 by dimartino
%-------------------------------------------------------------------
\subsection*{Acknowledgements}
\vskip 3 mm
 We are greatly indebted to our technical 
collaborators, to the members of the CERN-SL Division for the excellent 
performance of the LEP collider, and to the funding agencies for their
support in building and operating the DELPHI detector.\\
We acknowledge in particular the support of \\
Austrian Federal Ministry of Education, Science and Culture,
GZ 616.364/2-III/2a/98, \\
FNRS--FWO, Flanders Institute to encourage scientific and technological 
research in the industry (IWT), Federal Office for Scientific, Technical
and Cultural affairs (OSTC), Belgium,  \\
FINEP, CNPq, CAPES, FUJB and FAPERJ, Brazil, \\
Czech Ministry of Industry and Trade, GA CR 202/99/1362,\\
Commission of the European Communities (DG XII), \\
Direction des Sciences de la Mati$\grave{\mbox{\rm e}}$re, CEA, France, \\
Bundesministerium f$\ddot{\mbox{\rm u}}$r Bildung, Wissenschaft, Forschung 
und Technologie, Germany,\\
General Secretariat for Research and Technology, Greece, \\
National Science Foundation (NWO) and Foundation for Research on Matter (FOM),
The Netherlands, \\
Norwegian Research Council,  \\
State Committee for Scientific Research, Poland, SPUB-M/CERN/PO3/DZ296/2000,
SPUB-M/CERN/PO3/DZ297/2000 and 2P03B 104 19 and 2P03B 69 23(2002-2004)\\
JNICT--Junta Nacional de Investiga\c{c}\~{a}o Cient\'{\i}fica 
e Tecnol$\acute{\mbox{\rm o}}$gica, Portugal, \\
Vedecka grantova agentura MS SR, Slovakia, Nr. 95/5195/134, \\
Ministry of Science and Technology of the Republic of Slovenia, \\
CICYT, Spain, AEN99-0950 and AEN99-0761,  \\
The Swedish Natural Science Research Council,      \\
Particle Physics and Astronomy Research Council, UK, \\
Department of Energy, USA, DE-FG02-01ER41155, \\
EEC RTN contract HPRN-CT-00292-2002. 
%=========================================================================%

%=========================================================================%

\end{document}